\PassOptionsToPackage{table}{xcolor}
\documentclass[a4paper,fleqn]{cas-dc}
\definecolor{aliceblue}{rgb}{0.94, 0.97, 1.0}
\usepackage{rotating}
\usepackage{soul}

\usepackage{hyperref}
\usepackage{listings}

\usepackage{algorithmic}
\usepackage[ruled,vlined,linesnumbered]{algorithm2e}

\definecolor{codegreen}{rgb}{0,0.6,0}
\definecolor{codegray}{rgb}{0.5,0.5,0.5}
\definecolor{codepurple}{rgb}{0.58,0,0.82}
\definecolor{backcolour}{rgb}{0.97,0.97,0.97}
\definecolor{red}{RGB}{160,0,0}

\lstdefinestyle{mystyle}{
    backgroundcolor=\color{backcolour},   
    commentstyle=\color{codegreen},
    keywordstyle=\color{magenta},
    numberstyle=\tiny\color{codegray},
    stringstyle=\color{codepurple},
    basicstyle=\ttfamily\footnotesize,
    breakatwhitespace=false,         
    breaklines=true,  
    captionpos=b,                    
    keepspaces=true,                 
    numbers=left,                    
    numbersep=5pt,                  
    showspaces=false,                
    showstringspaces=false,
    showtabs=false,                  
    tabsize=2
}

\usepackage{array}
\usepackage{longtable}
\newcolumntype{N}{>{\centering\arraybackslash}m{.5in}}
\newcolumntype{G}{>{\centering\arraybackslash}m{2in}}
\usepackage{subfig}
\usepackage{mathtools}
\usepackage{balance}
\makeatletter
\usepackage{multirow}
\usepackage{colortbl}
\usepackage[table]{xcolor}
\usepackage[export]{adjustbox}
\def\BState{\State\hskip-\ALG@thistlm}
\makeatother
\usepackage{booktabs}
\usepackage{algorithmic}
\usepackage{graphicx}
\usepackage{textcomp}

\usepackage{csquotes}

\usepackage{tabularx}
\usepackage{tabularx,booktabs,ragged2e,float,adjustbox,makecell}
\usepackage[most]{tcolorbox}

\usepackage{circledsteps}
\pgfkeys{/csteps/fill color=aliceblue}
\usepackage{soul}
\hypersetup{
	colorlinks   = true, %Colours links instead of ugly boxes
	urlcolor     = blue, %Colour for external hyperlinks
	linkcolor    = blue, %Colour of internal links
	citecolor   = blue %Colour of citations
}
 \usepackage[numbers]{natbib}
\def\tsc#1{\csdef{#1}{\textsc{\lowercase{#1}}\xspace}}
\tsc{WGM}
\tsc{QE}
\tsc{EP}
\tsc{PMS}
\tsc{BEC}
\tsc{DE}
\newcommand{\tool}{{\it MuTAP}}
\begin{document}
\let\WriteBookmarks\relax
\def\floatpagepagefraction{1}
\def\textpagefraction{.001}

\shorttitle{Effective Test Generation Using Pre-trained Large Language Models and Mutation Testing}

\shortauthors{Arghavan MD et~al.}

\title [mode = title]{Effective Test Generation Using Pre-trained Large Language Models and Mutation Testing}                      
%\tnotemark[1,2]

%\tnotetext[1]{This document is the results of the research
  % project funded by the National Science Foundation.}

%\tnotetext[2]{The second title footnote which is a longer text matter
 %  to fill through the whole text width and overflow into
 %  another line in the footnotes area of the first page.}

\author[1]{Arghavan Moradi Dakhel}[orcid=0000-0003-1900-2850]

\cormark[1]


\ead{arghavan.moradi-dakhel@polymtl.ca}

%\ead[url]{www.cvr.cc, cvr@sayahna.org}

%\credit{Conceptualization of this study, Methodology, Software}

\affiliation[1]{organization={Department of Computer and Software Engineering},
    addressline={Polytechnique Montreal}, 
            city={Montreal},
            postcode={H3T 1J4},
            state={Quebec},
            country={Canada}}

\author[1]{Amin Nikanjam}
\ead{amin.nikanjam@polymtl.ca}
%third
\author[1]{Vahid Majdinasab}
\ead{vahid.majdinasab@polymtl.ca}
\author[1]{Foutse Khomh}
%[%
  % role=Co-ordinator,
  % suffix=Jr,
\ead{foutse.khomh@polymtl.ca}
\author[1]{Michel C. Desmarais}
\ead{michel.desmarais@polymtl.ca}

%\ead[URL]{www.sayahna.org}

%\credit{Data curation, Writing - Original draft preparation}

%\affiliation[2]{organization={Sayahna Foundation},
   % city={Jagathy},
    % citysep={}, % Uncomment if no comma needed between city and postcode
  %  postcode={695014}, 
   % state={Trivandrum},
   % country={India}}


\cortext[cor1]{Corresponding author}
%\cortext[cor2]{Principal corresponding author}

%\fntext[fn1]{This is the first author footnote. but is common to third
 % author as well.}
%\fntext[fn2]{Another author footnote, this is a very long footnote and
 % it should be a really long footnote. But this footnote is not yet
 % sufficiently long enough to make two lines of footnote text.}

%\nonumnote{This note has no numbers. In this work we demonstrate $a_b$
  %the formation Y\_1 of a new type of polariton on the interface
%  between a cuprous oxide slab and a polystyrene micro-sphere placed
 % on the slab.
 % }

\sloppy

\begin{abstract}
Context: One of the critical phases in the software development life cycle is software testing. Testing helps with identifying potential bugs and reducing maintenance costs. The goal of automated test generation tools is to ease the development of tests by suggesting efficient bug-revealing tests. Recently, researchers have leveraged Large Language Models (LLMs) of code to generate unit tests. While the code coverage of generated tests was usually assessed, the literature has acknowledged that the coverage is weakly correlated with the efficiency of tests in bug detection.\\
Objective: To improve over this limitation, in this paper, we introduce \tool{} (\textbf{Mu}tation \textbf{T}est case generation using \textbf{A}ugmented \textbf{P}rompt) for improving the effectiveness of test cases generated by LLMs in terms of revealing bugs by leveraging mutation testing.\\
Method: Our goal is achieved by augmenting prompts with surviving mutants, as those mutants highlight the limitations of test cases in detecting bugs. \tool{} is capable of generating effective test cases in the absence of natural language descriptions of the Program Under Test (PUTs). We employ different LLMs within \tool{} and evaluate their performance on different benchmarks.\\
Results: Our results show that our proposed method is able to detect up to 28\% more faulty human-written code snippets. Among these, 17\% remained undetected by both the current state-of-the-art fully-automated test generation tool (i.e., Pynguin) and zero-shot/few-shot learning approaches on LLMs. Furthermore, \tool{} achieves a Mutation Score (MS) of 93.57\%  on synthetic buggy code, outperforming all other approaches in our evaluation.\\
%\Michel{Could be more specific about the 79.  Which dataset, synthetic or real or both? And that means 79 more over what basis?---did it increase by 10\%, 100\%, ...?}
Conclusion: Our findings suggest that although LLMs can serve as a useful tool to generate test cases, they require specific post-processing steps to enhance the effectiveness of the generated test cases which may suffer from syntactic or functional errors and may be ineffective in detecting certain types of bugs and testing corner cases in \textit{PUT}s.
\end{abstract}

% \includegraphics{figs/grabs.pdf}

%\item Research highlights item 1
%\item Research highlights item 2
%\item Research highlights item 3

% Keywords
\maketitle
%%%%%%%%%%%%%%%%%%%%%%%%%%%%%%
\section{Introduction}
\label{sec:intro}

Testing is an important yet expensive step in the software development lifecycle. Generating effective tests is a time-consuming and tedious task for developers. Unit tests are essential as they form the basis of the test automation pyramid \cite{shore2021art,Siddiqui2021}. %In the hierarchy of software testing, unit tests are the smallest but essential components in modern software development \cite{shore2021art,Siddiqui2021}. 
Unit tests check if a function or a component works as expected in isolation. A unit test consists of two components: the first component is a set of test inputs for the Program Under Test (\textit{PUT}), while the second component is the test oracle that indicates the intended behavior (output) of the \textit{PUT} and is, therefore, capable of exposing bugs by verifying the correctness of the \textit{PUT} on test inputs~\cite{xie2006augmenting}. A test oracle can be in the format of assertions.
%A group of unit test cases is called a test set or test suite. 

%A group of unit tests is called test oracle. Test oracle can check the behavior of a function under normal and exceptional circumstances, called assertion oracle and exception oracle respectively.~\cite{fraser2010mutation, dinella2022toga}. 

The automatic generation of unit tests is an important topic in Software Engineering (SE). It aims to reduce developers' testing efforts. Developing good-quality unit tests can prevent bugs in software products. There are different tools for automatically generating unit tests and test suites that are either based on random test generators \cite{selakovic2018test,arteca2022nessie}, dynamic symbolic execution \cite{sen2005cute,godefroid2005dart}, or search-based approaches \cite{fraser2011evolutionary,fraser2011evosuite}. However, these techniques have some drawbacks and often generate tests with no assertion or too general assertions, or tests with assertions that cannot effectively assess the intended behavior of the \textit{PUT} \cite{panichella2020revisiting,palomba2016diffusion}.
%1) being less readable and hard to understand compared to manually developed test cases \cite{almasi2017industrial,grano2018empirical}, and 2) 
 %\Arghavan{Or we can say: As these methods mainly aim to optimize test sets based on code coverage, they face limitations to detect real bugs~\cite{almasi2017industrial}}. 

%\Arghavan{this is not a part of intro and it is for clarification only: They generate assertions, for example, pynguin generates assertions but generates them at the end after finalizing test cases by focusing on code coverage and since it cannot understand the functionality of the PUT, its test cases are less effective. EvoSuite uses some greedy techniques to remove redundant assertions.}

%They are used to generate test cases to check undesirable behavior of a code such as out of bound indexes, null references, etc. [ref].

%While the automatic test generator tools execute their test cases against the code under test, they have no understanding of the intention of the code.
%A production version of Codex is available as an extension on the Visual Studio Code development environment, named GitHub Copilot\footnote{\url{https://copilot.github.com/}}. Copilot 

Considering these shortcomings, researchers have recently been exploring the possibility of leveraging Machine Learning-based code synthesis techniques for generating better unit tests \cite{bareiss2022code, chen2022codet, lahiri2022interactive, schafer2023adaptive, lemieux2023codamosa}. Specifically, these approaches have been exploring the potential of Large Language Models (LLMs) with the transformer architecture, such as Codex~\cite{chen2021evaluating}, which has achieved good performance in automatic program synthesis \cite{brown2020language,chen2021evaluating,clement2020pymt5,feng2020codebert,MoradiJSS2023}. Among such efforts, Barei{\ss} et al. \cite{bareiss2022code} evaluate Codex's performance for test case generation by using a \textit{few-shot} learning approach. Their findings on a limited set of 18 Java methods show that their approach is comparable to feedback-directed test generation. ATHENATEST \cite{tufano2021unit} leveraged the BART transformer model \cite{lewis2019bart} after fine-tuning it on a set of real Java functions and their corresponding tests. They also reported achieving comparable coverage to EvoSuite \cite{fraser2011evosuite} after an assessment of five Java projects. Lemieux et al.  \cite{lemieux2023codamosa} proposed CODAMOSA which utilized test cases generated by Codex to improve search-based testing techniques, which consists of only the prefix (inputs) of a test case without any test oracles. Their reported results obtained on 27 Python projects show that CODAMOSA surpasses the baseline search-based technique, Pynguin~\cite{lukasczyk2023empirical} and Codex in terms of code coverage. Although the preliminary results of these studies and others \cite{tufano2022generating,schafer2023adaptive,chen2022codet, lahiri2022interactive}, are promising, none of these studies attempted to improve the bug detection capability of generated tests.
%\Foutse{the following sentence is too long!} 
Moreover, it has been acknowledged in the literature that while test coverage is a useful metric for evaluating the quality of tests, it is weakly correlated with the efficiency of tests in bug detection~\cite{CaiCovergae2005, gopinath2014code,hemmati2015effective}. %and is challenged as a measure of test effectiveness in revealing faults \cite{gopinath2014code,hemmati2015effective}.

%the effectiveness of generated tests guided by their bug-detecting capabilities.
%in real or synthetic buggy samples
%One such model is Codex~\cite{chen2021evaluating}; a GPT-3~\cite{brown2020language} based language model that shows impressive results in producing natural-looking code. It can generate code in different programming languages when provided with some context (called a prompt), such as comments, method names, or surrounding code. 

Mutation Testing (MT) is a white box testing technique to assess the capability of a test in revealing bugs. MT has been widely studied and successfully used in SE to assess the effectiveness of test cases \cite{Jia11, Papadakis19}. MT involves injecting \textit{artificial} changes based on \textit{real} faults into a \textit{PUT}, resulting in mutated versions of the \textit{PUT} known as mutants. The more a test case kills mutants, the more effective it is in identifying real bugs. The surviving mutants highlight the weaknesses of a test case and the ultimate goal is for the test cases to be able to detect all mutants, i.e., kill them. Mutants are not only useful for assessing the effectiveness of test cases but can also be used as a means for designing more effective test cases \cite{fraser2011evosuite}.

%None of the studies that leveraged LLMs to generate the test cases focused on mutation testing to improve the effectiveness of test cases.}
%the effectiveness of tests can then be scored by using the mutation score, the ratio of killed mutants by the total number of mutants.
%\Arghavan{Mutation testing helps in precisely this matter by suggesting not only where to test, but also what to check for. This is a very important aspect of mutation testing, and has hitherto not received attention for automated testing.}

%In this study, we present the first study on evaluating the competence of Codex and mutation testing, in generating an assertion oracle.

In this paper, we present the first study that leverages MT to enhance and evaluate the effectiveness of test cases generated by LLMs for Python programs in terms of fault revealing capabilities. Our approach aims to optimize test cases for bug detection rather than code coverage. Our proposed technique, \tool{}, employs an LLM as its main Component (LLMC) and starts by feeding a prompt to the LLMC in order to generate test cases. The initial prompt includes the \textit{PUT} and instructions for generating test cases by using \textit{zero-shot} and \textit{few-shot} learning. Next, \tool{} assesses the syntax of the generated test cases and re-prompts its LLMC to rectify any detected syntax issues. After fixing syntax errors, \tool{} proceeds to appraise the intended behavior of the generated test cases. This is achieved by comparing the output of the test oracles on certain test inputs to the expected return values of the PUT using the same test inputs, thereby correcting any unintended behavior in the test oracles. 

Subsequently, \tool{} applies MT to examine the effectiveness of test cases in killing mutants of \textit{PUT}s. As surviving mutants highlight the limitation of the generated test cases, \tool{} re-prompts its LLMC to generate new test cases for the \textit{PUT}s that have surviving mutants by augmenting the initial prompt with both initial test cases and the surviving mutants. \tool{} halts the process of augmenting the initial prompt when either the final test cases can effectively detect all mutants or there are no surviving mutants left that have not already been used to augment the initial prompt.
%\tool{} performs prompt augmentation on different surviving mutants until the final test cases can effectively detect all the mutants, or there are no surviving mutants left to augment with the initial prompt.

%\tool{} halts the process of augmenting the initial prompt when either the final test cases can effectively detect every mutant or there are no surviving mutants that can be augmented with the initial prompt.
%and re-prompting the LLMC.
% This is achieved
%
%In our proposed technique, \tool{}, the pre-trained LLM, using zero-shot and few-shot learning, is fed with prompts that include the implementation of a Program Under Test (PUT) and a request for a test in natural language using zero-shot and few-shot learning.
%Then the syntactic and semantic correctness of tests or generated is evaluated and we attempt to fix detected issues. Afterward, we assess the effectiveness of generated oracles in finding bugs in real and synthetic buggy samples. To generate synthetic buggy samples, when the real buggy samples were not available, we use mutation testing as non-detective mutants reveal shortcomings in the generated test suites. The outcome of these evaluations is used and then leveraged to improve the quality of generated tests by re-prompting the LLM. 
%\Arghavan{\tool{} employs Codex as its Large Language Model Component (LLMC). 

We employ two types of LLMs as the LLMC of \tool{}: \textit{Codex}, which is designed for code-related tasks, and \textit{llama-2-chat}, which is optimized for dialog use cases and versatile enough to accommodate a range of tasks, including programming. We evaluate \tool{} on both synthetic bugs of 164 \textit{PUT}s~\cite{chen2021evaluating} and $1710$ %real
buggy programs collected from a Python bug repairing benchmark~\cite{hu2019re}. Our results indicate that our proposed approach generates effective test cases with an average Mutation Score (MS, the ratio of killed mutants by the total number of mutants) of 93.57\%, outperforming both Pynguin (a state-of-the-art fully-automated test generation tool) and the conventional LLM-based zero-shot/few-shot learning techniques. 
%approach using LLM
%\Michel{I am not sure if using ``codes'' to refer to ``code segments'' is acceptable in English.  I believe when you want to refer to multiple segments of code, you can either say ``buggy code'', which could refer to a single segment or more, or write it in full, ``buggy code segments''.  If this is a practice you have seen in the literature, then forget this comment.}Arghavan{you are right, there is no 's'. I correct them over the text}
Furthermore, our approach detects up to 468 (28\%) more buggy code snippets written by humans than other comparable methods in our evaluation. Remarkably, it identifies 79 (17\%) buggy code snippets of humans that none of the other techniques are able to detect.
%\Michel{Similar comment to the abstract.  We need some idea of the relative improvement, not only absolute numbers.}
%(89.41\% vs 78.78\% on real).  (up to 10.63\% improvement  Also, it detects 374 . a state-of-the-art test suite generation for Python} %Our proposed approach to generate assertion oracles with Codex successfully generates assertion oracles for xx\Amin{to be updated} programming tasks that automatic test generation tools are not responsive to. Also, it detects up to yy\%\Amin{to be updated} more real buggy codes compared to automatic test generation tools. Moreover, we have compared \tool{} to Pynguin framework, a state-of-the-art test suite generation for Python \cite{lukasczyk2022pynguin}. 
%We also provide a method to present our suggestion on how to expand the capacity of this model (and similar ones) toward this direction.
To summarize, this paper makes the following contributions: 
\begin{itemize}
    \item We present the first study on leveraging MT to generate test cases with LLMs.
    
    \item We propose a prompt-based learning technique to improve the effectiveness of test cases by augmenting the prompts with both initial test cases and surviving mutants of a \textit{PUT}.
    
    \item We assess the effectiveness of generated tests in detecting bugs in real and synthetic buggy versions of \textit{PUT}s. %\Arghavan{is it a contribution? let's don't make bold our dataset and make them sensitive about the fact that they are not similar to real software projects.}
   
   %\item \st{We examine the usefulness of mutation testing to improve the genetest test by LLMs.}
   
    \item We make the proposed technique, \tool{}, publicly available online~\cite{rep-package} for other researchers/practitioners to replicate or build upon our work.
\end{itemize}

\textbf{The rest of this paper is organized as follows}. Section~\ref{sec:motiv_exmpl} introduces a motivating example. %of how \tool{} improves the effectiveness of test cases generated by LLMs. 
Section~\ref{sec:approach} describes the different steps of our approach. We present our experimental setup, research questions, and experimental results %to assess \tool{} in generating test cases 
in Section~\ref{sec:eval}. We discuss our findings and the potential use cases of our approach in Section~\ref{sec:disc}. Threats to the validity of our results %and limitations of our approach 
are reviewed in Section~\ref{sec:thvalid}. We briefly review the related works in Section~\ref{sec:related}. Finally, we conclude the paper in Section~\ref{sec:conclusion}; highlighting some avenues for future works.
%\Arghavan{These are exact sentences collected from other studies for inspiration on the importance of mutation testing: 4. this process (the paper means generating test oracles) requires a deep understanding of the source code and is a non-trivial task.5. we not only get guidance in where to test but also what to test for. (it means with test oracles)}

%%%%%%%%%%%%%%%%%%%%%%%%%%%%%%%%%%%%%%%%%%%%%%%

%%%%%%%%%%%%%%%%%%%%%%%%%%%%%%%%%%
\section{Motivating Example}
\label{sec:motiv_exmpl}
\begin{figure*}[htbp]
\centerline{\includegraphics[width=\textwidth]{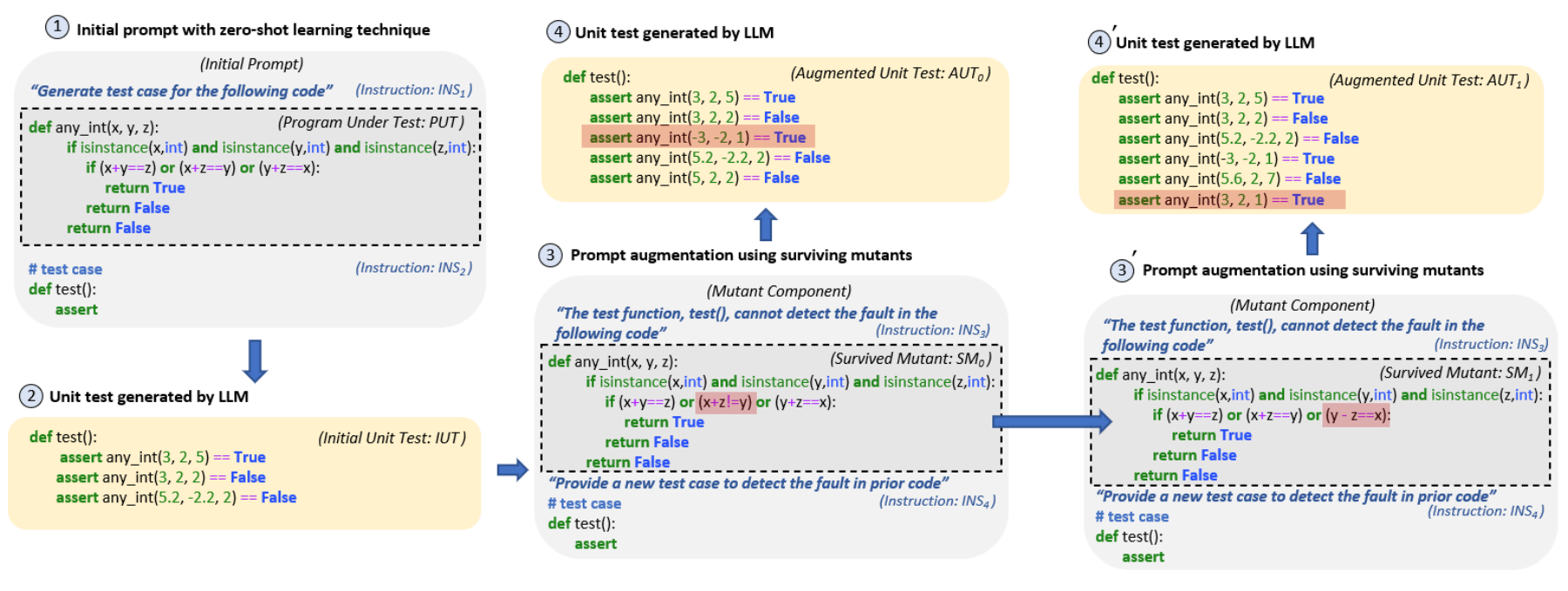}}
\caption{\textbf{Different steps of \tool{} on a PUT.} \Circled{2} is a set of test cases generated by the initial prompt \Circled{1} for $PUT$, and \Circled{4} is a set of test cases obtained after augmenting the initial prompt with the surviving mutant, \textit{SM$_0$}. \Circled{3}$^\prime$ shows the mutant component after updating with another surviving mutant of \textit{PUT$_0$} that we named \textit{SM$_{1}$.}}
%\Amin{one reviewer from ICSME asked for syntax/functional errors, did we include our reply in the paper: "The syntax error was uncompleted lines with missing parenthesis. The functional error was in the first and last assertion lines. Before the test repair, the expected output in the first line wrongly was “False” and in the last line was “True”."}}Arghavan: it is confusing adding them in the example since it needs first explanation on how to apply those steps
\label{fig:mutiv_smp}
\vspace{-1em}
\end{figure*}

In this section, we present an example  in Figure~\ref{fig:mutiv_smp} showing how our proposed approach generates effective test cases. Suppose we have 10 mutants $\{SM_{0}, SM_{1}, ..., SM_{9} \}$ for the Program Under Test, $PUT$ in Figure~\ref{fig:mutiv_smp}. 
The goal of our proposed technique, \tool{} (\textbf{Mu}tation \textbf{T}est case generation using \textbf{A}ugmented \textbf{P}rompt), is to generate effective test cases for $PUT$ 
in a way that ensures killing the maximum number of mutants. 

The function \textit{any\_int()} in Figure~\ref{fig:mutiv_smp} receives 3 inputs and returns \textit{True} if all 3 inputs are integers, also one of the inputs is equal to the sum of two others. Otherwise, it returns \textit{False}. In the first step, \tool{} uses the initial prompt, \Circled{1}, to run a query on the 
LLM Component (LLMC) and generates initial test cases for this 
Program Under Test (\textit{PUT}). %\Amin{we need to be consistent: PUT or FUT?}. 
The component \Circled{2} in Figure~\ref{fig:mutiv_smp} shows the initial test cases generated by LLMC after the refining step.
We named it Initial Unit Test, \textit{IUT}. In Section \ref{sec:approach}, we discuss the 
refining step (syntax and intended behavior fixing) of our approach in detail. 
The \textit{IUT} kills 6 out of 10 mutants of \textit{PUT}. The 4 remaining mutants reveal the weaknesses of the generated test, meaning that \textit{IUT} needs new test cases with assertion to kill the injected bugs in those 4 mutants.

To address this limitation and generate more effective test cases, \tool{} augments the initial prompt with two new components; the first 
one is the response of the model to the initial prompt after fixing its
syntax and intended behavior, \textit{IUT}, and the second one is the mutant component, \Circled{3} in Figure~\ref{fig:mutiv_smp}. \tool{} initiates the construction of the mutant component by using the first ``Survived Mutant'' of \textit{PUT} that we refer to as \textit{SM$_{0}$}. The red highlight in \textit{SM$_{0}$} shows the injected bug in $PUT$. The injected bug changes the second statement in the condition of the inner \textit{if} in $PUT$ in a way that the sum of the first and last input of function \textit{any\_int()} is not equal to the middle input anymore. Since there is no test case in \textit{IUT} to verify that its middle input, \textit{y}, is equal to the sum of its first and last inputs, \textit{x} and \textit{z}, \textit{IUT} is not able to kill this mutant.

\tool{} uses the concatenation of these three components: \Circled{1}, \Circled{2}, and \Circled{3} to re-prompt the LLMC. The \Circled{4} component in Figure~\ref{fig:mutiv_smp}, shows the new set of test cases generated by LLMC appended to \textit{IUT} after the refining step. We named it Augmented Unit Test, \textit{AUT$_0$}. The unit test has two more assertions compared to the \textit{IUT} and one of them, highlighted in red, kills the mutant, \textit{SM$_0$}.

\tool{} applied \textit{AUT$_0$} to the mutants of $PUT$ again. If there are any remaining surviving mutants, \tool{} iterates the augmentation process by updating the mutant component with another surviving mutant if it has not been used to augment the prompt previously. \tool{} utilizes each mutant individually because sometimes new test cases that address one mutant can also kill the remaining surviving mutants. Moreover, due to the limited length of the prompt and non-constant length of mutants, applying each surviving mutant separately is a more practical approach. Figure~\ref{fig:mutiv_smp} \Circled{3}$^\prime$ shows an example of how the mutant component is updated using another surviving mutant. We call this mutant \textit{SM$_1$}. Unit test, \Circled{4}$^\prime$, shows a new set of test cases including one assertion that detects \textit{SM$_1$}. 
\tool{} iterates the augmentation process until either the final test cases can kill all the mutants, or there are no surviving mutants left that have not already been used to augment the initial prompt. %In the final step, \tool{} applies a greedy approach to filter redundant assertions while maximizing the number of killed mutants. 

The final test cases generated by our proposed technique, \tool{}, kill 9 out of 10 mutants of this example, $PUT$, and it increases the MS for $PUT$ from $60\%$ (6 out of 10) to $90\%$ (9 out of 10). This result can be compared to the state-of-the-art automatic test generation tool for Python programming language~\cite{lukasczyk2023empirical}, Pynguin, which generates a test case for $PUT$ with only a $40\%$ MS. %\Foutse{it seems that you haven't defined the acronym MS at his point...it is only later in Section 3.3 that I see Mutation Score (MS)}.Arghavan{thanks, I added it into the abstract} 
This tool uses a search-based generation technique~\cite{arcuri2013parameter} and randomly mutates the test values within a test case to generate new test cases. The random nature of this method results in a low chance of generating a new test case that can kill the surviving mutants of \textit{PUT}. 

%However, by leveraging LLMs for test generation and incorporating mutation testing, \tool{} is able to produce more effective test cases than both the state-of-the-art tool and common zero-shot/few-shot learning methods using LLMs. %uses LLM as its core of test generation that is able to synthesize code and this competence helps \tool{} to generate more effective test cases.  %understands %\Amin{I'm afraid we can't claim that LLM understands the code!}the \textit{PUT} and its mutants. This competence helps \tool{} to generate more effective test oracles. 

%%%%%%%%%%%%%%%%%%%%%%%%%%%%%%%%%%%%%%%%%%%%%%%%%%%%%%%%%%%%%%%%%%%%%%%%%%%%%%%%%%%%%%
\section{Approach} \label{sec:approach}
\begin{figure*}[t]
\centerline{\includegraphics[width=0.95\textwidth]{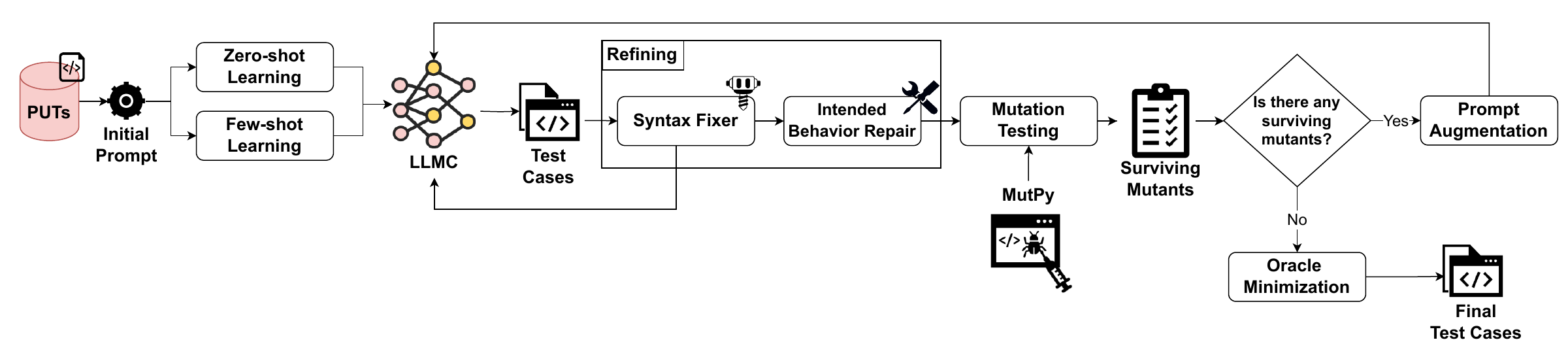}}
\caption{The proposed methodology for generating and evaluating tests using LLMs.}
\label{fig-methodology}
\vspace{-1em}
\end{figure*}

%\Foutse{the description of your approach here is too verbose and hard to follow....why don't you have an algorithm or a pseudo-code to describe the different steps of Mutap???}
In this section, we discuss the different steps of our approach. Figure \ref{fig-methodology} shows an overview of our proposed approach and Algorithm \ref{algo:algo1} presents the sequence of its different steps.

\begin{algorithm}[t]
\scriptsize
\SetKwFunction{GenerateInitialPrompt}{\textbf{GenerateInitialPrompt}}
 \SetKwFunction{CorrectAssertions}{CorrectAssertions}
 \SetKwFunction{MutationTesting}{\textbf{MutationTesting}}
 \SetKwFunction{OracleMinimization}{\textbf{OracleMinimization}}
 \SetKwFunction{Refining}{\textbf{Refining}}
 \SetKwFunction{AugmentingPrompt}{\textbf{AugmentingPrompt}}
 \SetKwFunction{LLMC}{\textbf{LLMC}}
 \KwIn{\textit{PUT}, 
\textit{LLMC}, \textit{initial\_prompt\_type}}
 \tcc{\textit{INS$_1$}, \textit{INS$_2$}, \textit{INS$_3$}, \textit{INS$_4$} and \textit{INS$_{fix}$} are global variable as natural language instructions for the prompts}
\KwOut{\textit{FUT} \tcp{Final Unit Test}} 

\SetAlgoLined
 \tcp{Initial Prompt}
 \textit{initial\_prompt} $\leftarrow$ \GenerateInitialPrompt(\textit{PUT}, \textit{initial\_prompt\_type})\\
 \textit{raw\_IUT} $\leftarrow$ \LLMC (\textit{initial\_prompt})\\

 \tcp{Syntax Fixer and Intended Behaviour Repair}
 \textit{IUT} $\leftarrow$ \Refining (\textit{raw\_IUT}, \textit{PUT})

 \tcp{Mutation Testing}
 \textit{MS}, \textit{surviving\_mutant} $\leftarrow$ \MutationTesting(\textit{PUT}, \textit{IUT})\\
 \uIf{\textit{MS} $< 100\%$}{
            \tcp{Prompt Augmentation}
            \textit{AUT} $\leftarrow$ \AugmentingPrompt(\textit{MS}, \textit{PUT}, \textit{initial\_prompt}, \textit{IUT}, \textit{surviving\_mutant})\\
            \tcp{Oracle Minimization}
            \textit{FUT} $\leftarrow$ \OracleMinimization(\textit{AUT})
    }\Else
    {
    \tcp{F: Oracle Minimization}
    \textit{FUT} $\leftarrow$ \OracleMinimization(\textit{IUT})
    }
 %$report \leftarrow{} \text{extractReportFromGraph(\textit{final})}$\\
 \Return \textit{FUT}
 \caption{\tool}
 \label{algo:algo1}
 %\vspace{-5pt}
\end{algorithm}

\subsection{Initial Prompt}\label{sec:initialPrompt}

LLMs are capable of performing those tasks that they are already trained for. Fine-tuning LLMs to perform a new task is computationally expensive. Also, there are LLMs such as Codex that show a very good performance in generating code but since they are closed-source, fine-tuning them for a new task is impossible.

Prompt-based learning~\cite{liu2023pre, zhang2022repairing} is an effective technique to adapt LLMs for new tasks. A prompt is a combination of natural language and/or programming language context and is used as an input to LLMs. There are studies showing that putting a natural language instruction as a hint (\textit{zero-shot learning})~\cite{lemieux2023codamosa,feng2020codebert, schafer2023adaptive} or several examples (\textit{few-shot learning})~\cite{nashidretrieval, brown2020language,ahmed2022few} in the prompt increases the capability of LLMs in performing a new task. 

\tool{} employs both \textit{zero-shot} and \textit{few-shot} learning to build the initial prompt and calls LLMC on them separately. This step is shown in Algorithm \ref{algo:algo2}. In more detail, we employ \textit{zero-shot} and \textit{few-shot} as follows: %\Amin{we may remove this, it makes misunderstanding I think:}The final test oracles generated by \tool{} can be based on \textit{zero-shot} or \textit{few-shot} strategy: 

%\begin{figure*}[t]
%\centerline{\includegraphics[width=0.75\textwidth]{new_mutant_1.png}}
%\caption{The proposed methodology to generate and evaluate tests using LLMs.}
%\label{fig:new_mut}
%\vspace{0.1em}
%\end{figure*}

\begin{itemize}
    \item \textit{zero-shot}: The initial prompt generated by \textit{zero-shot} technique contains three units, following the approach in~\cite{lemieux2023codamosa}. The component indicated by \Circled{1} in Figure~\ref{fig:mutiv_smp} shows an example of such a prompt. The first unit in this component is an instruction in a natural language named ${INS}_1$ and it clarifies the task by asking: \textit{``Generate test case for the following code''}. The second unit is the Program Under Test (\textit{$PUT$}) and the last unit is a set of instructions in a programming language named \textit{INS$_2$}. The \textit{INS$_2$} acts as a hint to indicate the desired output for LLMC. The concatenation of {$(INS_1, PUT, INS_2)$} %\Amin{$PUT_i$ -> $PUT$}
    builds the initial prompt for \textit{zero-shot} learning (Line 2 in Algorithm \ref{algo:algo2}). %\Amin{revised}%The \textit{PUT} is the only unit that changes in this type of initial prompt for each new task while \textit{$INS_1$} and \textit{$INS_2$} are the same for all tasks.
    
    \item \textit{few-shot}: Prompt generation based on \textit{few-shot} learning uses a chain of inputs and expected outputs related to the downstream task. There are different approaches for presenting the pair of input and output in the prompt. We follow the approach in~\cite{ahmed2022few} to build the initial prompt with \textit{few-shot} strategy in \tool{}. Considering the maximum possible length of tokens for LLMC (4k tokens in our study), \textit{few-shot} prompt includes two different demonstrative examples of a Method~(M) and a Unit Test~(UT) %{\footnotesize \texttt{$<$Method (M), Unit Test (UT)$>$}} and one \textit{PUT} 
    as follows (Line 5 in Algorithm \ref{algo:algo2}):\\ %\Amin{perhaps emphasize that we generate only test cases (assertions) not the whole unit test, one reviewer complained about this. We need to avoid implying being over-claimed.}:  Arghavan: last time we decided to call the group of test cases as a unit test. it is also called unit test in motivation example. for example, IUT stands for Initial Unit Test.

{\footnotesize \texttt{$<$code$>$M\_1$<$/code$>$\textbackslash n$<$test$>$UT\_1$<$/test$>$\textbackslash n $<$code$>$M\_2$<$/code$>$\textbackslash n$<$test$>$UT\_2$<$/test$>$\textbackslash n $<$code$>$PUT\_i$<$/code$>$\textbackslash n $<$test$>$}}

   % Same as \textit{zero-shot learning}, \textit{PUT} is the only unit that changes for each new task in this initial prompt and demonstrative examples stay the same.     
\end{itemize}
 
There is no natural language description of \textit{PUT} in the initial prompt since such descriptions may not always be available, and \tool{} relies on the ability of LLMC to synthesize code context. \tool{} calls the initial prompt, \textit{zero-shot} or \textit{few-shot}, on LLMC and then passes the inferred output to the next step (Line 2 in Algorithm \ref{algo:algo1}).

%\subsection{Large Language Model Component (LLMC)}\label{sec:LLMC}
%We utilize two different LLMs as the LLMC of \tool{}. The first one is OpenAI's Codex, designed specifically for code generation tasks. The second one is Meta's llama-2-chat, which has been iteratively refined using Reinforcement Learning with Human Feedback (RLHF) and is appropriate for dialog use cases. However, it is important to note that \tool{} is not limited to these two models, and its LLMC can be replaced with any other LLM as required.

\begin{algorithm}[t]
\scriptsize
\KwIn{\textit{PUT}, \textit{initial\_prompt\_type} }
\KwOut{\textit{initial\_prompt}}
\SetAlgoLined
  \uIf{\textit{initial\_prompt\_type} $==$ "zero-shot"}{
        \textit{initial\_prompt} $\leftarrow$ CONCAT(\textit{INS$_1$}, \textit{PUT}, \textit{INS$_2$})
        }
    \Else{
        \If{\textit{initial\_prompt\_type} $==$ "few-shot"}{
            \textit{initial\_prompt} $\leftarrow$ CONCAT(pair(\textit{M},\textit{UT}), \textit{PUT}) \tcp{M: Method, UT: Unit Test}
    }
    }
    \Return \textit{initial\_prompt}
 \caption{GenerateInitialPrompt}
 \label{algo:algo2}
 %\vspace{-5pt}
\end{algorithm}

\subsection{Refining} \label{sec:refine}
In this section, we describe the process of refining the generated test cases in \tool{} which includes fixing syntactical errors and intended behavior repair. The details are shown in Algorithm~\ref{algo:algo3}.
\subsubsection{Syntax Fixer}\label{sec:syntax-fix}
The test cases generated by LLMC may have syntax errors (missing brackets, uncompleted lines, etc.). Since \tool{} needs to execute the test function for investigation on MT and prompt augmentation, samples with syntax errors become inefficient. However, sometimes a small change in the output of LLMC can fix the syntactic error and convert it into an executable test case.

\tool{} uses the capability of its LLMC to fix syntax errors, similar to other studies \cite{zhang2022repairing, joshi2022repair}. To do so, LLMC is called on a new prompt to fix the syntax error in its own output (Procedure \textit{SyntaxFixer} in Algorithm \ref{algo:algo3}). The syntax fixing prompt 
consists of two parts. The first part is a natural language instruction, $INS_{fix}$, \textit{``Fix the syntax errors in the following code snippet''}, and the second 
part is the generated test function by LLMC on the initial prompt (Line 7-8 in Algorithm \ref{algo:algo3}). If the syntax error persists even after re-prompting the LLMC, \tool{} employs the Python parser to identify the erroneous line. It then retains the lines preceding the problematic line, ensuring they remain free of syntax errors (Line 13 in Algorithm \ref{algo:algo3}). 

\subsubsection{Intended Behavior Repair}\label{sec:semantic-fix}%\Amin{we may discuss but isn't it test correctness or functionality?}Fixing}
%\Amin{we may remove this sentence:}
%\Amin{we can remove the first sentence:}
Based on the initial prompt, LLMC generates different test cases that are serialized as an assertion oracle by calling the \textit{PUT} on certain inputs and comparing the returned output of \textit{PUT} with the expected output or ground truth, for example, \texttt{\{assert add (2,2) == 4\}}. However, it is possible for the LLMC to generate test cases that are asserting wrong return values. It means that for some test cases, LLMC does not generate the expected return output of the \textit{PUT}. The lack of a natural language description about the \textit{PUT} in the initial prompt could potentially lead to the generation of test cases that do not accurately reflect the intended behavior of the method.

%However, sometimes the LLMC generates test cases that have functional errors. The lack of any natural language description about \textit{PUT} in the initial prompt is a potential explanation of this observation. It means in some test cases, LLMC does not correctly generate the test inputs or the return output of the \textit{PUT}.

%Taking the assertion above as an example, suppose we have a \textit{PUT} named \verb|add| that receives two input arguments and returns their sum $\rightarrow$ \texttt{\{def add(input1,input2):\textbackslash  n  return (input1+input2)\}}. 
The assertion with wrong return values may fail on mutants, not because of detecting the bug, but because of the unintended behavior of the assertion. %The test cases with the functional error may fail on mutants as well because of such errors which 
These failures cause confusion about the effectiveness of test cases. So, this step of \tool{} aims at repairing the intended behavior of assertion oracles in the test cases (Procedure \textit{IntendedBehaviorFixer} in Algorithm \ref{algo:algo3}). 

For each assertion in the test, \tool{} runs the \textit{PUT} over the test inputs and compares the return output of \textit{PUT} with the asserting output. If the returned output of \textit{PUT} is the same as the asserting output in the oracle, then \tool{} considers it as an assertion oracle with the correct intended behavior. Otherwise, it repairs those assertions by replacing the asserting output with the expected output of \textit{PUT} (Line 22-27 in Algorithm \ref{algo:algo3}). \tool{} omits those assertions for which the input types failed on \textit{PUT}, for example, if \textit{PUT} expected a \verb|list| of integers but the test input is a \verb|string|. The final outcome of this step is named \textit{Initial Unit Test (IUT)} which is a set of test cases generated by LLMC after refinement as shown by \Circled{2} in Figure~\ref{fig:mutiv_smp}.

%The final outcome of this step is named \textit{Initial Unit Test (IUT)} which is a set of test cases generated by LLMC after fixing syntax and repairing their intended behavior. An example of \textit{IUT} is shown in \Circled{2} in Figure~\ref{fig:mutiv_smp}. This step also assesses the capability of LLMC to generate test cases for a \textit{PUT} with correct intended behavior even without a context describing the \textit{PUT}.

%This step also investigates the ability of LLMC in generating test cases for a \textit{PUT} that are functionally correct, without a context that describes the \textit{PUT}, by investigating the functional soundness of assertions in test cases.

\begin{algorithm}[t]
\SetKwFunction{IntendedBehaviorFixer}{IntendedBehaviorFixer}
\SetKwFunction{SyntaxFixer}{SyntaxFixer}
\SetKwFunction{SyntaxCheck}{SyntaxCheck}
\scriptsize
\KwIn{\textit{raw\_IUT}, \textit{PUT} }%\textit{$INS_fix$: ``Fix the syntax errors in the following code snippet''}
\KwOut{\textit{IUT} \tcp{ The Initial Unit Test after refining}}  
\SetAlgoLined
    
    \textit{syntax\_fixed\_IUT} $\leftarrow$ \SyntaxFixer(\textit{raw\_IUT})\\
     \textit{IUT} $\leftarrow$ \IntendedBehaviorFixer(\textit{syntax\_fixed\_IUT}, \textit{PUT})\\
    \Return \textit{IUT}\\
    \texttt{\\}
    \textit{\textbf{Procedure} SyntaxFixer(raw\_IUT)}\\
   \Indp \If{\textbf{not} AST.parse(\textit{raw\_IUT})}{
        \textit{syntax\_fixed\_prompt} $\leftarrow$ CONCAT (\textit{INS$_{fix}$}, \textit{raw\_IUT})\\
        \textit{syntax\_fixed\_IUT} $\leftarrow$ \LLMC (\textit{syntax\_fixed\_prompt})\\
        }
        \textit{syntax\_fixed\_IUT} $\leftarrow$ \SyntaxCheck (\textit{syntax\_fixed\_IUT})\\
     \Return \textit{syntax\_fixed\_IUT}\\
      \texttt{\\}
      
   \Indm \textit{\textbf{Procedure} SyntaxCheck(syntax\_fixed\_IUT)}
   
  \Indp \uIf{AST.parse(\textit{syntax\_fixed\_IUT})}{
        \Return \textit{syntax\_fixed\_IUT}}
    \Else{
        \Return \SyntaxCheck (\textit{syntax\_fixed\_IUT [:error\_line]})
    }

     \texttt{\\}
      
   \Indm \textit{\textbf{Procedure} IntendedBehaviorFixer(syntax\_fixed\_IUT, PUT)}

   \Indp \textit{fixed\_IUT} $\leftarrow \{\}$\\
    \ForEach{\textit{test\_case} $\in$ \textit{syntax\_fixed\_IUT}}{
          \textit{expected\_output} $\leftarrow$ \textit{PUT}(\textit{test\_case}.input)\\
          \uIf{\textit{expected\_output} $\neq$ \textit{test\_case}.output}{
                \textit{test\_case}.output $\leftarrow$ \textit{expected\_output}
          }
          \textit{fixed\_IUT}.append(\textit{test\_case})
          }
    \Return \textit{fixed\_IUT}
 \caption{Refining}
 \label{algo:algo3}
 %\vspace{-5pt}
\end{algorithm}

\subsection{Mutation Testing (MT)}\label{sec:mutation}
MT assesses the quality and effectiveness of test cases. 
Mutants are built by injecting artificial bugs into the \textit{PUT} to simulate defects. If test cases 
failed on a mutant, we consider
it as a killed mutant, otherwise, 
it survived, meaning that the test cases within the unit test are not able to detect it. The presence of surviving mutants highlights the shortcomings of test cases, suggesting the need to either add a new test case or improve an existing one.
The \textit{Mutation Score (MS)} represents the effectiveness of test cases by calculating the ratio of killed mutants out of all mutants of a \textit{PUT}.

Algorithm~\ref{algo:algo4} presents the details of this step. Inspired by~\cite{lukasczyk2022pynguin}, \tool{} uses MutPy \cite{mutpy} to generate different mutants for each \textit{PUT} and calculate \textit{MS} (Line 3-7 in Algorithm \ref{algo:algo4}). Executing test cases on each mutant involves performing some preliminary setups. For this purpose, \tool{} uses Python's built-in ``setuptools.find\_packages'' to locate and install the required packages, such as "math", "numPy", "pandas", "pytest", and others. Additionally, \tool{} implements setup functions that are responsible for creating temporary directories, which are utilized during the execution of the test cases on the mutants. After executing the test cases on the mutants and calculating the  \textit{MS}, \tool{} properly tears down the setup by removing the temporary directory. 

As shown on Line 5-9 in Algorithm \ref{algo:algo1}, if the \textit{MS} of a \textit{PUT} reaches \textit{100\%}, \tool{} passes test cases to the oracle minimization step (Subsection \ref{sec:greedy-opt}), otherwise, it collects the list of surviving mutants and transfers the mutants to the prompt augmentation step (Subsection \ref{sec:prompt-aug}).

\begin{algorithm}[t]
\scriptsize
\KwIn{\textit{PUT}, \textit{IUT}}
\KwOut{\textit{MS}, \textit{surviving\_mutant}}
\SetAlgoLined
     \textit{mutants} $\leftarrow$ MutPy(\textit{PUT})\\
     \textit{surviving\_mutant} $\leftarrow \{\}$\\
     \ForEach{\textit{MUT} $\in$ \textit{mutants}}{
		  \If{exec(\textit{MUT}, \textit{IUT})}{
			   surviving\_mutant.append(\textit{MUT})
            }
      }
     %\textit{MS} = (len(mutants)-len(survivng_mutant))/len(lst_mutants)
     \textit{MS} $\leftarrow (\#(\textit{mutants})-\#(\textit{surviving\_mutant}))/\#(\textit{mutants})$\\
    \Return \textit{MS}, \textit{surviving\_mutant}
 \caption{MutationTesting}
 \label{algo:algo4}
 %\vspace{-5pt}
\end{algorithm}

\subsection{Prompt Augmentation}\label{sec:prompt-aug}

Algorithm \ref{algo:algo5} shows the details of this step. %If there is any surviving mutant from the previous step, \tool{} augments the initial prompt, \textit{zero-shot} or \textit{few-shot}, by adding two new components. The first component is $IUT_i$\Amin{IUT\_i!}, the initial unit test generated by LLMC after syntax and functional correction. The second component is made of three units as shown by \Circled{3} in Figure~\ref{fig:mutiv_smp}. The first unit is an instruction in a natural language named \textit{$INS_3$} that clarifies the shortcoming of \textit{IUT} by \textit{``The test function, test(), cannot detect the fault in the following code''}. The second unit is the surviving mutants of the \textit{PUT}, named \textit{$SM_{ij}$}. The \textit{i} represents the identifier of \textit{PUT} and \textit{j} indicates the identifier of the surviving mutants. The last unit, \textit{$INS_4$} is an instruction in natural and programming language: the natural language context clarifies the task by asking to \textit{``Provide a new test case to detect the fault in prior code''} and the programming language context acts only as a hint to guide LLMC for generating the output.
If there is any surviving mutant from the previous step, \tool{} augments the initial prompt, \textit{zero-shot} or \textit{few-shot}, by adding four new components (Line 3 in Algorithm \ref{algo:algo5}). The first component is \textit{IUT}, the initial unit test generated by LLMC after refinement. The second component is an instruction in a natural language named \textit{INS$_3$} that clarifies the shortcoming of \textit{IUT} by \textit{``The test function, test(), cannot detect the fault in the following code''}. The third component is one of the surviving mutants of the \textit{PUT}, named \textit{SM}. The last component, \textit{INS$_4$} is an instruction in natural and programming language: the natural language context clarifies the task by asking to \textit{``Provide a new test case to detect the fault in prior code''} and the programming language context acts only as a hint to guide LLMC for generating the output. An example is shown by \Circled{3} in Figure~\ref{fig:mutiv_smp}.

%The concatenation of \textit{(\textit{Initial Prompt}, $IUT_i$, $INS_3$, $SM_{ij}$, $INS_4$)} builds the enhanced prompt. \tool{} reprompt LLMC and repeats the \textit{Syntax} and \textit{intended behavior} correctness steps on the generated output. Then, it appends new generated test cases to the $IUT_i$ that we call Augmented Unit Test  \textit{$AUT_{ij}$}. The \textit{$AUT_{ij}$} is passed to the \textit{MT} step. If the \textit{MS} is less than \textit{100\%}, then \tool{} repeats the \textit{Prompt Augmentation} step by updating the mutant component and replacing $SM_{ij}$ with another survived mutant $SM_{ij+1}$. An example of updating the mutant component in Figure~\ref{fig:mutiv_smp}, \Circled{3} is changed to \Circled{3}$^\prime$ by replacing $SM_{00}$ with  $SM_{01}$ and then the enhanced prompt is built by concatenating \textit{(\textit{Initial Prompt}, $IUT_i$, $INS_3$, $SM_{ij+1}$, $INS_4$)}. The \Circled{4}$^\prime$ indicates the generated test cases with LLMC after iterating the process on the next surviving mutant, we named it $AUT_{01}$.
%\Amin{revised, but must be updated according to the revised figure 1:}
\tool{} re-prompt LLMC and repeats the refining step on the generated output (Line 4-5 in Algorithm \ref{algo:algo5}). Then, it appends new generated test cases to the \textit{IUT} that we call Augmented Unit Test  (\textit{$AUT$}). The \textit{$AUT$} is passed to the \textit{MT} step (Line 7 in Algorithm \ref{algo:algo5}). \tool{} recursively repeats prompt augmentation till either the final test cases kill all the mutants ($\textit{MS} = \textit{100\%}$) or there is no surviving mutant that is not used in the augmentation process (Line 8 in Algorithm \ref{algo:algo5}). An example of updating the mutant component in Figure~\ref{fig:mutiv_smp}, \Circled{3} is changed to \Circled{3}$^\prime$ by replacing \textit{SM$_0$} with \textit{SM$_1$}. The \Circled{4}$^\prime$ indicates the generated test cases with LLMC after iterating the process on the next surviving mutant.

%The resulting test cases are then passed to the next step: the oracle minimization as shown in . 
%This means that \tool{} uses each surviving mutant at most once to enhance the prompt, and 
%%\Amin{exactly? Or “tries each mutant once”?} 

\begin{algorithm}[t]
\scriptsize
\KwIn{\textit{MS},  \textit{PUT}, \textit{initial\_prompt}, \textit{IUT}, \textit{surviving\_mutant}}
\KwOut{\textit{AUT} \tcp{Augmented Unit Test}}
\SetAlgoLined
      %\textit{AUT = IUT}\\
      
       \uIf{\textit{MS} $< 100\%$ or \textit{surviving\_mutant} $\neq$ \{\}}{
            
            \textit{SM} $\leftarrow$ \textit{surviving\_mutant}.pop()\\
		  \textit{augmented\_prompt} $\leftarrow$ CONCAT (\textit{initial\_prompt}, \textit{IUT}, \textit{INS$_3$}, \textit{SM}, \textit{INS$_4$})\\
            \textit{raw\_AUT} $\leftarrow$ \LLMC (\textit{augmented\_prompt})\\
            \textit{fixed\_AUT} $\leftarrow$ \Refining(\textit{raw\_AUT}, \textit{PUT})\\
            \textit{AUT} $\leftarrow$ \textit{IUT}.append(\textit{fixed\_AUT})\\ 
            \textit{MS} $\leftarrow$ \MutationTesting(\textit{AUT}, \textit{PUT})\\
	     \Return \AugmentingPrompt(\textit{MS},  \textit{PUT}, \textit{initial\_prompt}, 
      \textit{AUT}, \textit{surviving\_mutant})\\
       }
       \Else{\Return \textit{AUT}}
       %\Else{\uIf{\textit{AUT}}{\Return \textit{AUT}} \Else{\textit{IUT}} }
 \caption{AugmentingPrompt}
 \label{algo:algo5}
 %\vspace{-5pt}
\end{algorithm}

\subsection{Oracle Minimization} \label{sec:greedy-opt}
The test cases generated by the
LLMC usually consists of redundant assertions. Also, the augmentation process may add more redundant assertions to the final unit test. Presenting all of them (with redundancy) as the final output can cause confusion for developers. In the final step, similar to previous tools that generate mutation-driven test oracles~\cite{fraser2010mutation, fraser2011evosuite}, \tool{} minimizes the number of assertions by utilizing a Greedy technique to eliminate the redundant assertions that do not improve the MS. This step is presented in Algorithm \ref{algo:algo6}. \tool{} starts by tracking the number of mutants that each assertion kills and then chooses the test case containing the assertion that kills the maximum number of mutants.
This process is then repeated by adding the test cases containing the next assertions that detect the most mutants (Line 4-10 in Algorithm \ref{algo:algo6}). If adding this new assertion increases the MS, \tool{} keep the test case and its assertion. Otherwise, the test case will be discarded as redundant. %The output of this step named Final Unit Test (FUT).

\begin{algorithm}[t]
\scriptsize
\KwIn{\textit{PUT}, \textit{AUT}}
\KwOut{\textit{FUT}}
\SetAlgoLined
       \textit{MS\_old} $\leftarrow 0$\\
       \textit{FUT} $\leftarrow \{\}$\\ 
       
       \textit{sorted\_AUT} $\leftarrow \textit{\textbf{sort}(AUT)}$ \tcp{sort each test case in AUT based on the MS}
       
       \ForEach{\textit{test\_case} $\in$ \textit{sorted\_AUT}}{
           \textit{FUT}.append(\textit{test\_case})\\
           \textit{MS} $\leftarrow$ \MutationTesting(\textit{FUT}, \textit{PUT})\\
	   \uIf{\textit{MS} $>$ \textit{MS\_old}}{
                \textit{MS\_old} $\leftarrow$ \textit{MS}
                }
           \Else{
              \textit{FUT}.delete(\textit{test\_case})
              }
        }
    \Return \textit{FUT}

    % \texttt{\\}
      
 % \textit{\textbf{Procedure} sort(IUT, PUT)}
 % \SetAlgoLined
 \caption{OracleMinimization}
 
 \label{algo:algo6}
 %\vspace{-5pt}
\end{algorithm}
%%%%%%%%%%%%%%%%%%%%%%%%%%%%%%%%%%%%%%%%
\section{Evaluation} \label{sec:eval}

In this section, we describe the evaluations we designed and conducted to investigate the following research questions:

\begin{itemize}
    \item[\textbf{RQ1}]How effective are test cases generated by \tool{} in comparison to test cases generated by automatic test generation tools?
    \item[\textbf{RQ2}] How do the different parts of \tool{} perform?%\Amin{How do the different parts of this tool perform?}
    \item[\textbf{RQ3}] What is the performance of \tool{} for each mutation type? %How does \tool{} perform per mutation type?%?\Amin{or How does \tool{} perform per mutation type?}\Amin{I don’t think we assess correlation anyway!}
\end{itemize}

\subsection{Experimental Setup}
In this section, we present our experiment setup. Specifically, we describe the automatic test generation tool used to compare our results, clarify the LLMC of \tool{} and its setup,  and indicate the baselines and benchmark datasets used in our experiments. %, %experimental result, and describe %finally determine the benchmark datasets used in our experiments.

We conducted the experiment on the Cedar cluster of Compute Canada, which offers 32 cores CPU, 1TB storage, and one v100l GPU with 32GB GPU Memory, and on a system running Linux 5.15.0-69-generic with AMD FX(tm)-6300 Six-Cores CPU, 512GB storage, and 16GB Memory.

\subsubsection{\textbf{Experimental Parameters}}
We call the initial prompt, \textit{zero-shot} or \textit{few-shot}, on LLMC up to $10$ times and collect the outputs that meet two criteria as candidate test cases: the candidate should consist of two keywords of \textit{assert} and the \textit{function name} of \textit{PUT}. If after 10 runs, LLMC is not able to generate an output that contains those two keywords, we consider the task as a problematic task or a task for which \tool{} is not able to generate a test case.

Regarding the syntax fixing step, we run the syntax fixing prompt on the LLMC for up to 10 runs. If the syntax error remains unresolved even after 10 iterations, \tool{} employs the Python parser to locate the erroneous line. It then retains the lines preceding the buggy line, ensuring their freedom from syntax errors. If the removal of lines results in the absence of any remaining test cases (all test cases prove non-compilable), we classify the task as problematic.

\subsubsection{\textbf{Comparable Tool}}
Pynguin \cite{lukasczyk2022pynguin} is a well-known fully-automated test generation tool for a dynamically typed programming language such as Python. It uses different search-based algorithms toward satisfying code coverage criteria, i.e., branch coverage. Pynguin first takes a Python code (method, module, etc.) as input and collects its information such as variable types, method names, and dependencies. Then it uses one of the search-based test generation algorithms (MIO~\cite{arcuri2018test}, MOSA~\cite{panichella2015reformulating}, DynaMOSA~\cite{panichella2017automated}, etc.) to generate test cases. It randomly mutates (deletes, inserts, replaces) different values and statements within the test case to generate new test cases and executes them over the $PUT$ to ensure their correctness. Finally, it generates assertions for test cases using a MT engine~\cite{lukasczyk2022pynguin}.

For our experiments, we employ Pynguin 0.17.0. with the DynaMOSA~\cite{panichella2017automated}. 
%DynaMOSA is an evolutionary algorithm that uses branch coverage as its optimization problem. 
%DynaMOSA is a many-objective optimization technique designed to address the test case generation problem in the context of coverage testing which can dynamically select the coverage targets. 
According to the evaluation of Pynguin~\cite{panichella2017automated}, DynaMOSA shows the best performance compared to the other algorithm in generating test cases with this tool. We set the timeout of test generation to 600 seconds which is the default setting of the tool.

\subsubsection{\textbf{Large Language Model Component (LLMC)}}
We employ two different LLMs as the LLMC of \tool{}. The first one is OpenAI's Codex, designed specifically for code generation tasks~\cite{chen2021evaluating}. We use \textit{Code-davinci-002}, with a temperature of 0.8. The lower temperature causes less variation in the outputs of the model while the higher temperature increases the variation of output and then the chance of generating useful test cases over different iterations. The evaluation of CODAMOSA \cite{lemieux2023codamosa} shows that 0.8 is a reasonable temperature to generate useful test cases with Codex. 

The second LLM is Meta's \textit{llama-2-chat}, which has been iteratively refined using Reinforcement Learning with Human Feedback (RLHF) and is appropriate for dialog use cases~\cite{touvron2023llama}. Similar to Codex, we have configured the model's temperature to be 0.8. Furthermore, the model provides three distinct roles within the prompt:~\textit{system, user,} and \textit{assistant}. These roles serve the purpose of clarifying each component of the prompt to the model by assigning specific components to each role. Different combinations of these roles can be utilized in each prompt to tailor the interaction with the model according to the specific requirements~\cite{touvron2023llama}.%\Amin{any citation to be added?}. 

In our experiments, the role of the \textit{system} is defined as \textit{\{You are a Python coding assistant. Always answer with Python code.\}}, for all types of prompts, including \textit{zero-shot}, \textit{few-shot}, and \textit{augmented} prompts. To handle the \textit{zero-shot} prompt, we only set the \textit{user's} role content to be a concatenation of \textit{(INS$_1$, PUT$_i$, INS$_2$)}. For the \textit{few-shot} prompt, we define the content of the \textit{assistant} role as a set of demonstrative examples of Method~(M) and Unit Test~(UT), while the \textit{user} role content is set to \textit{PUT$_i$}. As for the \textit{augmented} prompt, its various components are set up as follows:\\
{\footnotesize \texttt{\{user: Initial Prompt,\\ assistant: \textit{IUT},\\ user: concat(\textit{INS$_3$}, \textit{SM$_{i}$}, \textit{INS$_4$})\}}}

%\tool{} uses OpenAI’s Codex model, specifically \textit{code-cushman-001}, with a temperature of 0.8 as its LLMC. The lower temperature causes less variation in the outputs of the model while the higher temperature increases the variation of output and then the chance of generating useful test cases over different iterations. The evaluation of CODAMOSA \cite{lemieux2023codamosa} shows that 0.8 is a reasonable temperature to generate useful test cases with Codex.

For both LLMs, the maximum number of generated tokens is set to 250 for generating test cases and 20 tokens for syntax fixing, based on previous studies on similar tasks~\cite{lemieux2023codamosa, shrivastava2022repository}. The stop word is defined as \textit{quote (``)} for \textit{zero-shot} and as \textit{$</test>$} for \textit{few-shot} prompt. For the rest of the hyperparameters, we keep the model's default values.
%\Michel{verify quote that is not closed}

To avoid overfitting on the benchmarks data, \tool{} repeats all prompts on Codex or llama-2-chat for up to 10 runs. If after 10 runs, the requirement for generating test cases is not satisfied, \tool{} considers it as a problematic or unsolved task. %as explained in Subsection \ref{sec:initialPrompt}.

It is important to note that \tool{} is not limited to these two models, and its LLMC can be replaced with any other LLM as required.
\subsubsection{\textbf{Baselines}}
In addition to Pynguin, we propose two baselines for each LLM to evaluate our proposed method, \tool{}.\\ 
\textit{\textbf{Before-refining:}} The first baseline is the output of the initial prompt on LLMC (Codex or llama-2-chat), without fixing syntax errors or repairing the intended behavior. Since assertions with unintended return values can fail on mutants or buggy code and present invalid effectiveness, we omit those assertions in this baseline to avoid this side effect. If the output of the model has syntax errors, we consider it as a wrong test and consequently consider the task as a problematic or unsolved task.\\
\textit{\textbf{After-refining:}} The second baseline is the output of the initial prompt on LLMC (Codex or llama-2-chat), after applying the following steps: \textit{Refining} (Subsection \ref{sec:refine}) and \textit{Oracle Minimization} (Subsection \ref{sec:greedy-opt}).  

\subsubsection{\textbf{Mutant Generator}}
To apply MT, we need to generate different mutant versions of a \textit{PUT} by injecting bugs into its different lines. For this purpose, we use \textit{MutPy} version 2.0 \cite{mutpy}. \textit{MutPy} is a MT tool for code in Python 3.3+. It benefits from different mutation operators to generate the mutants. The list of mutation operators used in our experiment with corresponding examples is shown in Table \ref{mutation-operators}. \textit{MutPy} injects one operator at a time to generate the mutant if the operator is applicable on \textit{PUT}.

%\newcolumntype{g}{>{\columncolor{lightgray}}c}
\begin{table*}[htbp]
    \caption{List of the mutation operators in our experiments used by $MutPy$ sorted by alphabetical order.}
    \centering
    \rowcolors{2}{gray!20}{white}
    \resizebox{0.75\textwidth}{!}{ 
    \begin{tabular}{lll}
     \hline
     \rowcolor{gray!45}
        \multicolumn{1}{c }{\textbf{\textbf{Operator}}} &  \multicolumn{1}{c }{\textbf{\textbf{Example}}}  &  \multicolumn{1}{c }{\textbf{\textbf{Mutant}}}  \\ 
         \toprule
        %\addlinespace
        %\hline
        AOD - arithmetic operator deletion &  result.append(numbers[-1]) &  result.append(numbers[1])  \\ 
         \addlinespace[3pt]
         %\hline
        AOR - arithmetic operator replacement & return number \% 1.0 & return number * 1.0  \\ 
       \addlinespace[3pt]
        %\hline
        ASR - assignment operator replacement & current\_depth += 1 & current\_depth -= 1  \\ 
        \addlinespace[3pt]
        %\hline
        BCR - break continue replacement & if i \% j != 0: break & if i \% j != 0: continue  \\ 
        %\hline
        \addlinespace[3pt]
        COD - conditional operator deletion & if not string:  return ' ' & if string:  return ' '  \\
        \addlinespace[3pt]
        %\hline
        COI - conditional operator insertion & if balance $<$ 0: return True & if (not balance $<$ 0): return True  \\ 
        %\hline
        \addlinespace[3pt]
        EHD - exception handler deletion & except: pass & except: raise  \\ 
        %\hline
        \addlinespace[3pt]
        EXS - exception swallowing & except: return False & except: pass  \\ 
        %\hline
        \addlinespace[3pt]
        LCR - logical connector replacement & if s[-1] == 'y' or s[-1] == 'Y': & if s[-1] == 'y' and s[-1] == 'Y':  \\ 
        %\hline
        \addlinespace[3pt]
        ROR - relational operator replacement & if c[n] $<=$ 1: & if c[n] $>=$ 1:  \\ 
        %\hline
        \addlinespace[3pt]
        SIR - slice index remove & l[::3] = sorted(l[::3]) & l[::3] = sorted(l[:]) \\ 
        \bottomrule
    \end{tabular}
    }
    \label{mutation-operators}
    %\vspace{1mm}
\end{table*}

\subsubsection{\textbf{Benchmark Datasets}} To conduct our experiments, we use two different benchmarks. The first one is \textit{HumanEval} \cite{chen2021evaluating} which is a benchmark to evaluate LLMs that generate code. It has 164 human-written programming problems at easy to medium levels. Each problem has different attributes such as descriptions and reference solutions. We use the reference solution of each task as a \textit{PUT}. 

The second one, \textit{Refactory} \cite{hu2019re}, is a benchmark for Python bug repairing \cite{re-bug-repair}. It has 1710 buggy students' submissions for 5 assignments of a Python programming course. Each assignment has a correct reference solution that we use as \textit{PUT}. The advantage of this dataset is buggy code snippets generated by humans that give us the opportunity to evaluate test cases generated by \tool{} on real bugs and compare it with Pynguin and our baselines.

%There are different advantages in these two datasets:\\
%Created by human. It reduces the chance of memorization problems by Copilot.\\
%Easy to medium level. Copilot is mostly responsive to easy/medium programming problems.
%Reference Solution. For each problem, we have a reference solution that can be beneficial to check the correctness of each unit test.\\
%Human assertion oracle. Each problem has a set of input/output as an assertion oracle generated by humans to test the functionality of the code in normal behavior which can act as a baseline for our study.
%%%%%%%%%%%%%%%%%%%%%%%%%%%%%%%%%%%%%%%%
\subsection{Experimental Results}
In this section, we discuss our findings for each RQ.
\subsubsection{\textbf{RQ1: How effective are test cases generated by \tool{} in comparison to test cases generated by automatic test generation tools?}}
Since our study focuses on MT to improve the effectiveness of test cases, we compare \tool{} with Pynguin and our baselines in terms of MS, number of killed mutants, and number of \textit{PUT} with $100\%$ MS. It is worth mentioning that we only consider \textit{PUT}s with correct test cases to calculate the average MS for each method. For this reason, we report the total number of killed mutants and the total number of \textit{PUT}s with $100\%$ MS for a fair comparison.

\newcolumntype{g}{>{\columncolor{gray!20}}c}

\begin{table*}[htbp]
\caption{Evaluation result of test cases generated by \tool{} and other methods on \textit{synthetic} buggy programs.}
\begin{center}
\resizebox{0.85\textwidth}{!}{%
%\rowcolors{2}{gray!20}{white}
    \begin{tabular}{c c c c c c c c}
    \hline
    \rowcolor{gray!45}
   \textbf{Prompt} & \textbf{Model}& \textbf{Method} &
   \textbf{\# Test Cases (avg)} &  \makecell{\textbf{\# Problematic PUT}\\ \textbf{(out of 164)}} & \textbf{MS (\%)}  & \makecell{\textbf{\# Killed Mut} \\ \textbf{(out of 1260)}} & \makecell{\textbf{Task MS=100\%} \\ \textbf{(out of 164)}}\\
        \toprule
        - & -& Pynguin & 1.5 (min=1, max=4) & 31 & 65.94\% & 649 & 28.22\% (46)  \\ 
        \addlinespace [3pt]
        \hline
         && Before-refining & 1.5 (min=1, max=3) & 73 & 72.15\% & 296 & 11.04\% (18)  \\
         
         %\rowcolor{white}
         \makecell{\textbf{Zero-shot}} &\makecell{\textbf{Codex}} & after-refining & 2.1 (min=1, max=3) & 30 & 76.82\% & 749 & 24.54\% (40)  \\ 
        
         && \tool{} & \textbf{2.5 (min=1, max=4)} & \textbf{30} & \textbf{89.13\%} & \textbf{869} & \textbf{41.72\% (68)}  \\ 
        \hline
        %\cmidrule(lr){2-8}
        %\rowcolor{white}
        && Before-refining & 1.2 (min=1, max=3)& 68 & 62.60\% & 318 & 17.79\% (29)  \\

         \makecell{\textbf{Zero-shot}} &\makecell{\textbf{llama2-chat}} & After-refining &  2.2 (min=1, max=5) & 0 & 84.04\% & 1059 & 53.98\% (88)  \\ 
        %\rowcolor{white}
         && \tool{} & \textbf{2.5 (min=1, max=5)} & \textbf{0} & \textbf{91.98\%} & \textbf{1159} & \textbf{68.09\% (111)}  \\ 
        \hline
        %\addlinespace [3pt]
        %\rowcolor{white}
        && Before-refining & 1.5 (min=1, max=3) & 39 & 72.68\% & 508 & 15.95\% (26)  \\ 
        %\addlinespace [3pt]
         \textbf{ Few-shot } &\makecell{\textbf{Codex}} & After-refining & 2.2 (min=1, max=3) & 27 & 82.73\% & 829 & 34.97\% (57)  \\
         \addlinespace [3pt]
         %\rowcolor{white}
         && \tool{} & \textbf{2.6 (min=1, max=7)} & \textbf{27} & \textbf{92.02\%} & \textbf{922} & \textbf{49.69\%} \textbf{(81)} \\ 
          \hline
         && Before-refining &  1.5 (min=1, max=3) & 60 & 64.51\% & 325 & 22.69\% (37)  \\ 
        %\addlinespace [3pt]
       
         \textbf{ Few-shot } &\makecell{\textbf{llama2-chat}} & After-refining & 2.5 (min=1, max=5) & 0 & 85.16\% & 1073 & 75.05\% (93)  \\
         \addlinespace [3pt]
         %\rowcolor{white}
         && \tool{} & \textbf{2.6 (min=1, max=7)} & \textbf{0} & \textbf{93.57\%} & \textbf{1179} & \textbf{69.93\%} \textbf{(114)} \\\bottomrule
       
    \end{tabular}
    }
\end{center}
\label{HumanEval-overall}
\vspace{-2em}
\end{table*}

Table~\ref{HumanEval-overall} shows the obtained results for the \textit{HumanEval} benchmark. Prior to syntax fixing and intended behavior repair (\textit{before-refining}), the test cases generated by Codex and llama-2-chat are incorrect for 73 and 68 (out of 164) \textit{PUT}s, respectively, when using the \textit{zero-shot} initial prompt. However, they manage to kill 295 and 318 mutants (out of 1260), respectively.

The initial prompt has a more pronounced impact on the output of Codex compared to llama-2-chat. Switching the initial prompt to \textit{few-shot} decreases the number of \textit{PUT}s without test cases to 39, while also raising the number of killed mutants to 508 when using Codex as LLMC. On the other hand, when using llama-2-chat, the number of \textit{PUT}s without test cases reduces to 60, and the number of killed mutants increases from 318 to 325. This difference in performance could be attributed to llama-2-chat being more suitable for dialog prompts, and using a prompt with a pair of demonstrative input and output, devoid of natural language context, does not improve the model's performance significantly.%as LLMC for \tool{}

In contrast, Pynguin, as the state-of-the-art automatic test generation tool, outperforms the output of both LLMs, before-refining, by killing 649 mutants and failing to generate test cases for 31 tasks.

After applying the post-processing steps of syntax fixing and intended behavior repair, \tool{} with both LLMs perform better than Pynguin in terms of killing more mutants. Notably, when using both \textit{zero-shot} and \textit{few-shot} prompts, llama-2-chat is able to generate correct test cases for all \textit{PUT}s, after-refining. However, their effectiveness in terms of killing mutants is measured at 84.04\% and 85.16\% with the \textit{zero-shot} and \textit{few-shot} prompts, respectively.

On the other hand, the MS of test cases generated by Codex after refining is 76.82\% and 82.73\% with the \textit{zero-shot} and \textit{few-shot} prompts, respectively. Despite this improvement, Codex still fails in generating correct test cases for 30 (with \textit{zero-shot}) and 27 (with \textit{few-shot}) \textit{PUT}s after refining.

\tool{}, enhances the effectiveness of test cases generated by both LLMs, Codex and llama-2-chat, achieving MS of 89.13\% and 91.98\% with the \textit{zero-shot} prompt, and an MS of 92.02\% and 93.57\% with the \textit{few-shot} prompt, respectively. Particularly, \tool{} with the \textit{few-shot} prompt when using llama-2-chat as its LLMC manages to kill 1179 mutants out of 1260 and generates test cases with MS=100\% for up to 70\% of \textit{PUT}s, demonstrating a remarkable improvement in the effectiveness of test cases compared to the Pynguin with 649 killed mutants and 28.22\% \textit{PUT}s with MS=100\%.

%The average number of test cases for each \textit{PUT} equals $1.5$ with a minimum of $1$ and a maximum of $4$ test cases for both Pynguin and \tool{} with the \textit{zero-shot} initial prompt (after the oracle minimization step). When \tool{} uses the \textit{few-shot} initial prompt, the average number of its test cases increases to $2.6$ with a minimum of $1$ and a maximum of $7$.

Table~\ref{StudentEval-overall} that shows the results on the humans’ real buggy programs from \textit{Refactory} benchmark confirms our findings on \textit{HumanEval}. To evaluate \tool{} on real buggy code, we apply the following steps. First, we generate the mutants of each \textit{PUT} in this dataset. Second, we conduct the prompt augmentation process and finalize the test cases for each \textit{PUT}. Then, we apply test cases generated by \tool{} on students' buggy code in \textit{Refactory}, followed by test cases generated by Pynguin and LLMs \textit{After-refining}, to assess the effectiveness of test cases generated by different methods in detecting buggy code.

\tool{} with \textit{few-shot} learning while using llama-2-chat as its LLMC identifies $468$ more buggy code compared to Pynguin (with an MS of $94.91\%$  vs. $67.54\%$) and $111$ more buggy code compared to \textit{After-refining} (with an MS of $94.91\%$ vs. $82.51\%$). Furthermore, \tool{} discovers $79$ buggy code that were not detected by either Pynguin or llama-2-chat's test cases~\textit{After-refining} process.  When using Codex, \tool{} detects $73$ buggy code that were missed by both Pynguin and Codex's test cases \textit{After-refining} stage. Moreover, \tool{} excels in generating more effective test cases, with an average of $2.6$ test cases after applying greedy optimization.

Overall, \tool{} using both llama-2-chat and Codex demonstrates better performance compared to Pynguin in terms of killing mutants and detecting buggy code. The effectiveness of these test cases in detecting defects is improved through post-processing steps of refining and prompt augmentation.

%Our results indicate that Codex as an LLM, can generate test cases that are more effective than the ones generated by automatic test generation tools, however, the effectiveness of these test cases in detecting defects can be improved through post-processing repair and prompt augmentation.
%Our results suggest the test oracles generated by LLMs require post-processing to be more effective in detecting buggy programs and augmenting prompts with survived mutatans
%\Amin{Also what is the message of our results to practitioners? What can we propose for future work? To extend, improve, any suggestion, fine-tuning? I already wrote something at the end of abstract. }Arghavan: I added takeaway note instead of finding in the box due to space limitation

\begin{table}[htbp]
\caption{Evaluation results on \textit{real} buggy programs.}
    \centering
    \resizebox{\columnwidth}{!}{% 
    %\rowcolors{2}{gray!20}{white}
    \begin{tabular}{ c g g g g}
    \hline
        \rowcolor{gray!45}
        \textbf{Prompt} & \textbf{Model} &\textbf{Method} & \textbf{\# Test Cases (avg)} &  \makecell {\textbf{Bug} \textbf{Detected}\\ \textbf{(out of 1710)}}\\
        \hline
        \addlinespace [3pt]
        \rowcolor{white}
       - &- & Pynguin & 1.25 (min=1, max=4)  & 67.54\%  (1155)   \\ 
        \addlinespace [3pt]
        \hline
        \rowcolor{white}
         & & After-refining & 1.2 (min=1, max=2)  & 79.87\% (1356)  \\
         \addlinespace [3pt]
        \rowcolor{white}
        \multirow{-2}{*}{\textbf{Zero-shot}}& \multirow{-2}{*}{\textbf{Codex}} & \tool{} & \textbf{1.6 (min=1, max=3)} & \textbf{84.03\%}  \textbf{(1437)} \\
        \addlinespace [3pt]
        \hline
        \rowcolor{white}
         & & After-refining & 1.2 (min=1, max=3)  & 86.43\% (1478) \\
         \addlinespace [3pt]
        \rowcolor{white}
        \multirow{-2}{*}{\textbf{Zero-shot}}& \multirow{-2}{*}{\textbf{llama-2-chat}} & \tool{} & \textbf{2.2 (min=1, max=4)} & \textbf{93.22\%}  \textbf{(1594)} \\
        \addlinespace [3pt] 
        \hline
        \rowcolor{white}
         & & After-refining & 1.6 (min=1, max=3) & 82.51\%  (1411) \\
         \addlinespace [3pt]
        \rowcolor{white}
        \multirow{-2}{*}{\textbf{Few-shot}} & \multirow{-2}{*}{\textbf{Codex}}&\tool{} & \textbf{2.2 (min=1, max=4)} &  \textbf{89.41\%}  \textbf{(1529)}  \\ \addlinespace [3pt]\hline

        \rowcolor{white}
         & & After-refining & 2.1 (min=1, max=4) & 88.42\%  (1512) \\
         \addlinespace [3pt]
        \rowcolor{white}
        \multirow{-2}{*}{\textbf{Few-shot}} &\multirow{-2}{*}{\textbf{llama-2-chat}} &\tool{} & \textbf{2.2 (min=1, max=4)} &  \textbf{94.91\%}  \textbf{(1623)}  \\ \addlinespace [3pt]\hline 
    \end{tabular}
    }
\label{StudentEval-overall}
%\vspace{-2em}
\end{table}

\begin{tcolorbox}[colback=gray!8,colframe=gray!40!black]
\textbf{Finding 1:} \tool{} generates more effective test cases compared to Pynguin and conventional \textit{zero-shot} and \textit{few-shot} learning on LLM. The number of \tool{}'s test cases is not much greater than the output of other methods after minimization. Additionally, LLM with dialog setup performs better on the augmented prompt. In conclusion, the effectiveness of LLM-generated test cases can be enhanced through prompt augmentation using surviving mutants and post-processing refinement.%\Amin{we can compare performance of codex and llama too!} 

%\tool{} detects 374 more buggy codes compared to the Pynguin and 118 more buggy codes compared to the few-shot learning approach with the \textit{Codex}.
%\textbf{Challenges:} \Amin{is there any challenge we faced to add here?}
\end{tcolorbox}

%%%%%%%%%%%%%%%%%%%%%%%%%%%%%%%%%%%%%%%%%%%%%
\subsubsection{\textbf{RQ2: How do the different parts of \tool{} perform?}}
%In the term of the initial prompt type, as it is shown in Table~\ref{HumanEval-overall} and Table~\ref{StudentEval-overall}, using the few-shot learning technique causes generating correct test oracles for a more number of PUTs and detecting more buggy codes of students. 
%with a minimum of 3 and a maximum of 34 improvements compared to the zero-shot technique. Also, it increases the MS and the number of killed mutants. 
\textit{\textbf{Syntax Fixer:}} On average, the percentage of test cases with syntax errors is $38.98\%$ and $26.48\%$ when using the \textit{zero-shot} and \textit{few-shot} prompts, respectively, with Codex. When employing llama-2-chat, this percentage is $33.85\%$ and $26.32\%$ with the \textit{zero-shot} and \textit{few-shot} prompts, respectively. 

When considering syntax errors, three factors contribute to decreasing them in the output of LLMs. The first factor is the type of initial prompt. As shown in Table~\ref{HumanEval-Syntax} on the \textit{HumanEval} benchmark, \textit{few-shot} learning results in fewer syntax errors in the output of both LLMs. Specifically, when using Codex, the percentage of syntax errors decreases from $44.79\%$ to $29.03\%$ after-refining, and for \tool{}, it decreases from $33.17\%$ to $23.93\%$. With llama-2-chat as the LLMC, the percentage of syntax errors decreases from $38.03\%$ to $26.99\%$ after refining, and from $29.66\%$ to $25.64\%$ for \tool{}.

\begin{table*}[htbp]
\caption{Syntax error fixing of test cases. The syntax Error Rate shows the ratio of unit tests with syntax errors.}
    \centering
    \resizebox{0.85\textwidth}{!}{%
    \begin{tabular}{c c c c c c c c}
    \hline
        %\multicolumn{2}{|c|} 
        \rowcolor{gray!45}
        \textbf{Model}&\textbf{Method}&
        \textbf{Prompt} & \textbf{\# Run (avg)}& \textbf{Syntax Error Rate} & \textbf{Fixed by Model} & \textbf{Fixed by Omitting Lines}  \\
        \toprule

        \rowcolor{white}
        ~ & &Zero-shot & 9.1 & 44.79\% & 16.44\% & 60.27\%   \\ 
        \rowcolor{white}
         & \multirow{-2}{*}{\textbf{After-refining}} &Few-shot & 9.5 & 29.03\% & 12.96\% & 83.33\%  \\

%\cmidrule{2-6}
%\hline
         %\rowcolor{white}
       %  ~ & \textbf{Average Total} & 34.36\% &	14.63\% & 72.44\% & 12.93\%  \\
       % \hline
        %\hline
        \cmidrule(lr){2-7}
        \rowcolor{white}
        \multirow{-2}{*}{\textbf{Codex}} & ~ &Zero-shot & 9.7& 33.17\% & 16.18\% & 79.41\%   \\ 
        \rowcolor{white}
       & \multirow{-2}{*}{\textbf{\textit{MuTAP}}} & Few-shot & 9.5&23.93\% & 12.82\% & 84.62\% \\
        \hline
      %  ~ & \textbf{Average Total} & 31.10\%	& 14.57\%	& 81.37\% & 4.06\% \\
        \hline

        \rowcolor{white}
        ~ & & Zero-shot  & 7.1 & 38.03\% & 30.64\% & 63.86\%   \\ 
        \rowcolor{white}
        &\multirow{-2}{*}{\textbf{After-refining}} &Few-shot & 6.8& 26.99\% & 31.81\% & 57.96\% &  \\ 
        
%\cmidrule{2-6}
%\hline
         %\rowcolor{white}
       %  ~ & \textbf{Average Total} & 34.36\% &	14.63\% & 72.44\% & 12.93\%  \\
       % \hline
        %\hline
        \cmidrule(lr){2-7}
        \rowcolor{white}
       \multirow{-2}{*}{\textbf{llama2-chat}} & &Zero-shot &6.9& 29.66\% & 32.17\% & 61.05\%  \\ 
        \rowcolor{white}
       & \multirow{-2}{*}{\textbf{\textit{MuTAP}}} & Few-shot &6.8& 25.64\% & 32.45\% &60.40\%  \\
        \hline
      %  ~ & \textbf{Average Total} & 31.10\%	& 14.57\%	& 81.37\% & 4.06\% \\

    \end{tabular}
    }
\label{HumanEval-Syntax}
\vspace{-1em}
\end{table*}

The second impactful factor, which is also the primary factor, is the \textit{Syntax Fixing} component. As shown in Table~\ref{HumanEval-Syntax}, when using Codex, this component in \tool{} on average fixes $14.5\%$ of syntax errors by utilizing the LLMC and addresses $81.37\%$ of syntax errors by omitting the lines causing the errors. On the other hand, when using llama-2-chat as the LLMC of \tool{}, the \textit{Syntax Fixing} component, on average, resolves $32.31\%$ of syntax errors through re-prompting the LLMC, and $60.73\%$ of the errors by omitting the problematic lines.

The final factor contributing to the improvement of syntax errors in test cases is the prompt augmentation process in \tool{}. By augmenting the prompt with \textit{IUT}, the occurrence of syntax errors in the output of Codex with the \textit{zero-shot} technique decreases from $44.79\%$ to $33.17\%$. Similarly, with llama-2-chat and the \textit{zero-shot} prompt, the percentage of syntax errors reduces from $38.03\%$ to $29.66\%$. Augmenting the prompt with \textit{IUT} provides illustrative examples of test cases and serves a similar purpose to the demonstrative examples in the \textit{few-shot} learning prompt, effectively reducing syntax errors in the output of LLMs.

%For \textit{Refactory} dataset, since it was for 5 PUTs (5 assignments), we didn't report it in a separate table. 
Our finding on the \textit{Refactory} benchmark shows \tool{} generates test cases with syntax errors in only one \textit{PUT} (out of $5$) using Codex and \textit{zero-shot} learning.  Moreover, none of those syntax errors could be fixed by re-prompting LLMC. On the other hand, for both initial prompt types, syntax errors decrease to zero using llama-2-chat.

\textit{\textbf{Intended Behavior Repair:}} In the case of repairing intended behavior, two distinct factors contribute to reducing the error rate in assertion oracles. As shown in Table~\ref{HumanEval-Semantic}, the \textit{Intended Behavior Repair} step, when using Codex as the LLMC, on average, fixes $83.98\%$ and $89.86\%$ of incorrect behaviors in the \textit{after-refining} and \tool{}, respectively. When utilizing llama-2-chat, this step repairs $84.35\%$ and $95.96\%$ of unintended behavior in the \textit{after-refining} and \tool{}, respectively.

In addition to the \textit{Intended Behavior Repair} step, the prompt augmentation step in \tool{} significantly reduces the occurrence of unintended behavior in test cases. For instance, when using Codex with a \textit{zero-shot} prompt, the assertions with unintended behavior, such as wrong return values, decrease from $63.63\%$ to $19.38\%$. Similarly, with llama-2-chat and using \textit{few-shot} prompt, the assertions with unintended behavior decrease from $63.25\%$ to $10.75\%$. The reason behind this improvement could be attributed to the usage of \textit{IUT}s (Initial Unit Tests) in \tool{} for augmenting the initial prompt. These \textit{IUT}s already represent the intended behavior of the $PUT$, thereby assisting the LLM in suggesting test cases with less unintended behavior (i.e., fewer wrong return values). Also, on the \textit{Refactory} benchmark, \tool{} repaired all assertions with incorrect behavior on the output of augmented prompts.

Unlike syntax errors, the prompt type does not significantly help with unintended behavior in assertions. %\Amin{the prompt type does not significantly improve/help with unintended behavior?}.
The combination of the \textit{Intended Behavior Repair} step and the prompt augmentation process improves the effectiveness %\Amin{effectiveness?otherwise accuracy in what?} 
of test cases, ensuring that they align with the intended behavior of \textit{PUT}.

\begin{figure}[t]
\centerline{\includegraphics[width=0.475\textwidth]{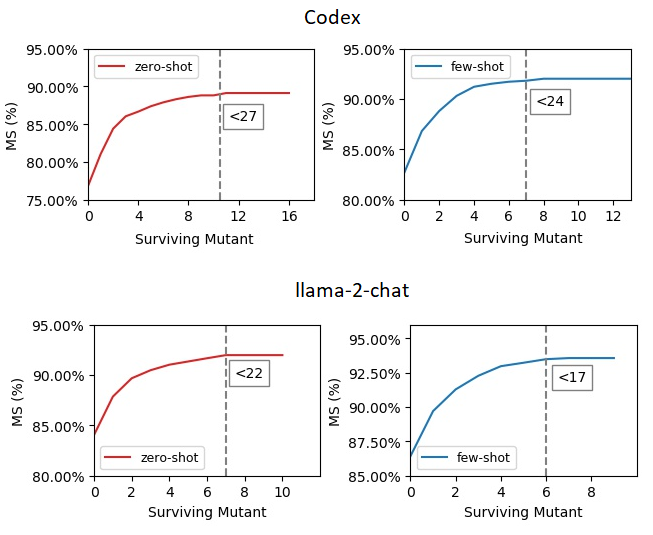}}
\caption{The impact of utilizing surviving mutants in different random orders on the MS for different LLMs. Each data point represents the average MS for all PUTs across five different runs, wherein the surviving mutants were randomly selected for the prompt augmentation process.}%\Amin{in the label of x-axis, "surviving mutant"}}
\label{fig:survivedmut}
\vspace{-1em}
\end{figure}

\textit{\textbf{Surviving Mutants Representation:}} We also investigated the impact of surviving mutants' order on MS during prompt augmentation. Figure~\ref{fig:survivedmut} illustrates the effect of augmenting the prompt with a random order of surviving mutants over $5$ runs for all \textit{PUT}s. For this comparison, we randomly selected one of the surviving mutants of each \textit{PUT} with $\textit{MS}<100\%$ and utilized it to augment the initial prompt. We then calculated the average MS for all \textit{PUT}s. Subsequently, we randomly chose the second surviving mutant for the remaining \textit{PUT}s with $\textrm{MS}<100\%$ (if any), repeated the augmentation process, and calculated the average MS for all \textit{PUT}s again. We continue repeating this process until either there are no more \textit{PUT}s with $\textrm{MS}<100\%$  or no more surviving mutant that is not utilized in the argumentation process.

As shown in Figure~\ref{fig:survivedmut}, each data point represents the average MS for all \textit{PUT}s over 5 runs of a random selection of surviving mutants. Notably, more than $90\%$ of the MS is achieved by using only half of the surviving mutants, and the improvement in MS stalls after a certain repetition of the augmentation step in different LLMs. For example, when using Codex as LLMC, in \textit{zero-shot} learning, the MS stops improving even though, on average, $27$ surviving mutants (out of $226$) are not utilized in the prompt augmentation step. Similarly, in \textit{few-shot} learning, this number is equal to $24$ (out of $106$). %\Amin{to me, codex performs a bit better, no?}Arghavan: the total number of surviving mutants with llama is less than codex. the percentage of surviving mutants that didn't contribute in MS for both model are pretty close. plus surviving mutant for llama is out of all tasks but codex has difficulty for some tasks.

%\vspace{-25pt}
\begin{table}[t]
\caption{Evaluation results of \textit{Intended Behavior Repair}. The Assertion Error Rate shows the ratio of assertions with wrong behavior.}
    \centering
    \resizebox{\columnwidth}{!}{% 
    \begin{tabular}{g g g g g g}
    \hline
       %\multicolumn{2}{|c|}{\multirow{2}{*}
       \rowcolor{gray!45}
       \textbf{Model}&\textbf{Method}&
       \textbf{Prompt} & 
       \makecell{\textbf{Assertion} \\ \textbf{Error Rate}} & \textbf{Repaired} &
       %\textbf{Fixed} & \multirow{2}{*}
       \textbf{Not Repaired}\\ 
        %\multicolumn{2}{|c|}{} & \textbf{Error} & \textbf{Semantic Error} & \\
        \hline
        \rowcolor{white}
      ~ & ~ & \textbf{Zero-shot} & 63.63\% & 82.21\% & 17.79\%  \\
        \rowcolor{white}
        &\multirow{-2}{*}{After-refining} & \textbf{Few-shot} & 62.84\% & 85.75\% & 14.25\%  \\
        %\hline
       % ~ & \textbf{Average Total} & 63.23\% & 83.98\% & 16.02\%  \\ 
        %\hline
        \cmidrule(lr){2-6}
        \rowcolor{white}
         \multirow{-2}{*}{\textbf{Codex}}&~& \textbf{Zero-shot} & 19.38\% & 89.71\% & 10.29\%  \\
        \rowcolor{white}
        &\multirow{-2}{*}{\textbf{\tool{}}}  & \textbf{Few-shot} & 18.36\% & 90.00\% & 10.71\%  \\
        \hline
        % & \textbf{Average Total} & 18.87\% & 89.86\% & 10.50\% \\
        \hline

         \rowcolor{white}
      ~ & ~ & \textbf{Zero-shot} & 60.27\%	& 81.80\% & 18.19\%  \\
        \rowcolor{white}
        &\multirow{-2}{*}{After-refining} & \textbf{Few-shot} & 63.25\% &	86.90\% &	13.09\% \\
        %\hline
       % ~ & \textbf{Average Total} & 63.23\% & 83.98\% & 16.02\%  \\ 
        %\hline
        \cmidrule(lr){2-6}
        \rowcolor{white}
         \multirow{-2}{*}{\textbf{llama-2-chat}}&~& \textbf{Zero-shot} & 23.40\%	& 94.06\% &	5.94\% \\
        \rowcolor{white}
        &\multirow{-2}{*}{\textbf{\tool{}}}  & \textbf{Few-shot} & 10.75\%	& 94.91\%	& 5.09\% \\
        \hline
        % & \textbf{Average Total} & 18.87\% & 89.86\% & 10.50\% \\
    \end{tabular}
    }
\label{HumanEval-Semantic}
%\vspace{-2em}
\end{table}
%\vspace{-15pt}

Our results for RQ2 demonstrate that test cases generated by LLMs, regardless of the prompt type, require post-processing, such as syntax correction or intended behavior repair, in order to function properly and detect bugs effectively. Also, the order of surviving mutants to augment the prompt does not significantly impact the MS gain.

%that each part of \tool{} positively  impacts the effectivness of test oracles generated by 
%\Amin{again what is the message of our results to practitioners? future work, suggestion?}
%\vspace{-3pt}
\begin{tcolorbox}[colback=gray!8,colframe=gray!40!black]
\textbf{Finding 2:} The \textit{Syntax Fixing} and \textit{Intended Behavior Repair} fix up to $95.94\%$ and $89.86\%$ of syntax and functional errors in test cases, respectively. The prompt augmentation in \tool{} decreases the unintended behavior in the output of LLMs significantly ($44.36\%$ using Codex and $52.5\%$ using llama-2-chat). Furthermore, only a small number of mutants (up to $27$) do not contribute to the improvement of MS.
\end{tcolorbox}
%%%%%%%%%%%%%%%%%%%%%%%%%%%%%%%%%%
\subsubsection{\textbf{RQ3: What is the performance of \tool{} for each mutation type?}}%\Foutse{see my previous comment on the formultion of thid rq}}\Amin{this is the new formulation}} 
In this RQ, we evaluate the performance of \tool{} in different mutant types. We report the total number and number of killed mutants by each method on the \textit{HumanEval} benchmark in Table~\ref{mutations-detail}. The performance of all techniques per mutant type is reported to help the comparison as well. The total number of mutants in each type is different for each method since the number of problematic \textit{PUT}s is not the same for all methods. The MS for each type/method indicates the ratio of killed mutants out of the total number of mutants in that type. There are some mutant types that are more common (more samples in those types) such as \textit{AOR}, \textit{COI}, and \textit{ROR} (an example for each mutant type is shown in Table~\ref{mutation-operators}). The number of mutants in each type depends on the \textit{PUT}. For example, in the \textit{HumanEval}, there are few \textit{PUT}s with exception handling. Consequently, there are few mutants in the \textit{EHD}.

In general, \tool{} shows better or similar performance in all mutant types compared to Pynguin and the output of LLMs \textit{After-refining} of both LLMs. Considering \textit{ASR} as an example, \tool{} shows the highest performance on this mutant type among all methods. For example, test cases generated by Pynguin identified 45 mutants in this category  while test cases generated by \tool{} using llama-2-chat and the \textit{few-shot} prompt identified 79 mutants in this category (out of 84).

For one of the mutant types, \textit{BCR}, which is a rare type in our benchmarks, \tool{} and \textit{After-refining} with both \textit{zero-shot} and \textit{few-shot} initial prompt, along with using Codex, show the same performance. However, when employing llama-2-chat, \tool{} outperforms the others by killing more mutants of this type. For another rare type of mutant in our dataset, \textit{EHD}, it is noteworthy that Codex, despite using both initial prompt types and the augmentation process, fails to generate test cases to detect the two mutants present in this category. In contrast, \tool{} with the \textit{few-shot} initial prompt and llama-2-chat successfully killed all of the mutants in this category.

%Our result shows that there are mutant types such 
%\Amin{similarly what is the message to practitioners? future work, suggestion?}

\begin{tcolorbox}[colback=gray!8,colframe=gray!40!black]
\textbf{Finding 3:} The test cases generated by \tool{} are equally or more effective in killing different types of mutants compared to those generated by Pynguin and the baseline method. Also, using an LLM with dialog setup can increase the number of killing mutants in different mutant types while applying prompt augmentation.%\Amin{and for LLMC, llama performs better?}
\end{tcolorbox}
%\textit{\textbf{Answer to RQ3:} Test oracles generated by \tool{} for different types of mutants outperform (or comparably perform as) the test oracles generated by Pynguin and both baselines.}

%\begin{table*}[!ht]
%\caption{StudentEval Dataset (Syntax Error Fixing).}
  %  \centering
  %  \begin{tabular}{|l|l|l|l|l|l|}
  %  \hline
  %      ~ & ~ & avg \# of run & syntax error (out of 5) & fixed with model & fixed by cutting lines  \\ \hline
   %     output of original prompt & zeroshot & 10 & 2 & 0 & 2  \\ \hline
    %    ~ & fewshot & - & - & - & -  \\ \hline
    %    output of refined prompt & zeroshot & 10 & 1 & 0 & 1  \\ \hline
     %   ~ & fewshot & 0 & - & - & - \\ \hline
  %  \end{tabular}
%\end{table*}

\begin{table*}[!ht]
    \caption{Evaluation of killed mutants for each type of injected operator into PUTs.}
    %\rowcolors{2}{white}{gray!20}
    \centering
     \resizebox{1\textwidth}{!}{
    \begin{tabular}{c|c c|c c|c c|c c|c c|c c|c c|c c|c c}
    \hline
       %\rowcolor{gray!45}
         \multirow{3}{*}{} & \multicolumn{2}{c|}{\multirow{2}{*}{\textbf{Pynguin}}} & \multicolumn{8}{c|}{\textbf{Zero-shot}} & \multicolumn{8}{c}{\textbf{Few-shot}}  \\
        %\hline
          %\cmidrule(lr){2-3}
         %\cmidrule(ll){4-15}
         \cline{4-19}
         %\cmidrule(lr){10-15}
        & \multicolumn{2}{c|}{} & \multicolumn{4}{c|}{\textbf{Codex}} & \multicolumn{4}{c|}{\textbf{llama-2-chat}} & \multicolumn{4}{c|}{\textbf{Codex}} & \multicolumn{4}{c}{\textbf{llama-2-chat}} \\
         \cline{4-19}
         %\rowcolor{gray!20}
         & \multicolumn{2}{c|}{} & \multicolumn{2}{c|}{\textbf{After-refining}} & \multicolumn{2}{c|}{\textbf{\textit{MuTAP}}} & \multicolumn{2}{c|}{\textbf{After-refining}} & \multicolumn{2}{c|}{\textbf{\textit{MuTAP}}} & \multicolumn{2}{c|}{\textbf{After-refining}} & \multicolumn{2}{c}{\textbf{\textit{MuTAP}}} & \multicolumn{2}{c|}{\textbf{After-refining}} & \multicolumn{2}{c}{\textbf{\textit{MuTAP}}}\\
         
          %\cmidrule(lr){1-1}
         %\cmidrule(lr){2-3}
         %\cmidrule(lr){4-5}
        % \cmidrule(lr){6-7}
         %\cmidrule(lr){8-9}
        % \cmidrule(lr){10-11}
        % \cmidrule(lr){12-13}
        % \cmidrule(lr){14-15}
         \hline
         \rowcolor{gray!45}
         \textbf{Type} & \textbf{killed} & \textbf{total} & \textbf{killed} & \textbf{total} & \textbf{killed} & \textbf{total} & \textbf{killed} & \textbf{total} & \textbf{killed} & \textbf{total} & \textbf{killed} & \textbf{total} & \textbf{killed} & \textbf{total}  & \textbf{killed} & \textbf{total}  & \textbf{killed} & \textbf{total}  \\ 
         
      \hline

        AOD	&13 (39.39\%)	&33& 20 (62.50\%) &	32	&28  (87.50\%) &	32	&36 (80.00\%)	&45	&39 (86.67\%)&	45&	27 (79.41\%)	&34	&32 (94.12\%)	&34	&37 (82.22\%)	&45&	40 (88.89\%)&	45\\ 
        \rowcolor{gray!20}
        AOR	&248 (67.39\%)	&368&	274 (74.66\%)&	367	&336 (91.55\%)	&367	&390 (87.05\%)	&448	&410 (91.52\%)	&448	&290 (77.33\%) &	375&	347 (92.53\%)	&375	&394 (87.95\%)&	448	&417 (93.08\%)&	448\\
        
        ASR	&45 (60.00\%)	&75&	56 (74.67\%)&	75&	60 (80.00\%)&	75	&74 (88.10\%)&	84	&79 (94.05\%)&	84&	57 (76.00\%)	&75&	64 (85.33\%)&	75	&75 (89.29\%)&	84	&79 (94.05\%)	&84 \\
        
        \rowcolor{gray!20}
        BCR&	2 (40.00\%)&	5&	2 (40.00\%)&	5	&2 (40.00\%)	&5	&5 (55.56\%)&	9	&5 (55.56\%)	&9	& 2 (40.00\%)	&5	&2 (40.00\%)	&5	&5 (55.56\%)&	9&	6 (66.67\%)	&9 \\
        
        COD&	8 (53.33\%)&	15	&12 (80.00\%)	&15&	15 (100.00\%)	&15&	15 (68.18\%)	&22&	16 (72.73\%)&	22&	15 (88.24\%)&	17	&17 (100.00\%)&	17&	15 (68.18\%)	&22&	17 (77.27\%)	&22\\ 
        \rowcolor{gray!20}
        COI&	130 (81.76\%)&	159	&145 (91.19\%)&	159	&154 (96.86\%)&	159	&194 (85.46\%)&	227&	216 (95.15\%)	&227	&161 (96.99\%)	&166&	164 (98.80\%)&	166&	200 (88.11\%)&	227&	218 (96.04\%)&	227 \\ 
        
        EHD	&1 (100.00\%)&	1&	0 (0.00\%)	&0	&0 (0.00\%)&	0&	1 (50.00\%)	&2&	2 (100.00\%)&	2	&0 (0.00\%)	&1&	1 (100.00\%)&	1&	1 (50.00\%)	&2	&2 (100.00\%)&	2\\ 
        \rowcolor{gray!20}
       EXS&	0 (0.00\%)&	0&	0 (0.00\%)	&1	&1 (100.00\%)&	1	&1 (100.00\%)&	1&	1 (100.00\%)&	1	&0 (0.00\%)&	1	&1 (100.00\%)&	1	&1 (100.00\%)&	1&	1 (100.00\%)&	1 \\ 
        
       LCR	&14 (45.16\%)&	31&	22 (70.97\%)	&31&	23 (74.19\%)	&31&	30 (69.77\%)&	43&	37 (86.05\%)&	43&	24 (72.73\%)&	33	&27 (81.82\%)&	33	&32 (74.42\%)&	43	&39 (90.70\%)	&43\\ 
        \rowcolor{gray!20}
        ROR&	174 (66.67\%)	&261&	200 (76.92\%)&	260&	227 (87.31\%)&	260&	281 (84.13\%)&	334&	316 (94.61\%)&	334&	227 (86.97\%)&	261&	239 (91.57\%)&	261	&282 (84.43\%)&	334	&320 (95.81\%)	&334 \\ 
        
       SIR&	10 (33.33\%)&	30	&18 (60.00\%)&	30&	23 (76.67\%)&	30	&32 (71.11\%)	&45	&38 (84.44\%)	&45&	26 (76.47\%)&	34	&28 (82.35\%)	&34&	31 (68.89\%)	&45&	40 (88.89\%)&	45 \\ 
        \hline
        \rowcolor{gray!20}
        Total&	645 (65.95\%)	&978&	749 (76.82\%)&	975&	869 (89.13\%)&	975&	1059 (84.05\%)&	1260&	1159 (91.98\%)&	1260&	829 (82.73\%)	&1002&	922 (92.02\%)	&1002&	1073 (85.16\%)	&1260&	1179 (93.57\%)&	1260\\ \hline
    \end{tabular}
    }
 \label{mutations-detail}
 \vspace{-0.5em}
\end{table*}

%%%%%%%%%%%%%%%%%%%%%%%%%%%%%%%%%%%%%%%%%%%%%%%%%%%%%%%%

%%%%%%%%%%%%%%%%%%%%%%%%%%%%%%%%%%%%%%%
\section{Discussion} \label{sec:disc}
\subsection{\textbf{Automatic Test Case Generation}}%\Amin{Automatic test case generation}} 
\tool{} leverages the code synthesis capabilities of LLMs and employs prompt-based learning to assist developers in generating effective test cases without the need for the computationally expensive fine-tuning of LLMs.

LLMs are able to generate test cases that are more effective than those generated by Pynguin in terms of revealing bugs. Listing~\ref{fig:pynguin_smp} shows a sample test case generated by Pynguin for the \textit{PUT} of our motivation example in Section~\ref{sec:motiv_exmpl}. While Pynguin generates test inputs as random integers and mutates those values to generate new test cases, LLMs produce test cases that are more natural-looking and correlated with input/output type and the functionality of the \textit{PUT}. However, test cases generated by LLMs require post-processing to become more effective in detecting bugs. Our results show that augmenting the prompt with surviving mutants and refining test cases (syntax and intended behavior) helps LLMs generate more effective test cases in terms of fault detection. %ng bugs. 

Developers can use \tool{} to generate effective test cases in terms of fault detection, with the help of LLMs. Additionally, \tool{} can be integrated into the test generation component of the GitHub Copilot lab \cite{copilotlab} to suggest more effective test cases for developers. Since the mutants can be generated automatically, prompt augmentation can be applied without human engagement. 

%Fine-tuning an LLM on a dataset of test cases, with a focus on generating more effective tests for bug detection, and subsequently comparing it with \tool{}, is left for future research.

\begin{lstlisting}[language=python, caption={A sample test case generated by Pynguin for the PUT in the motivation example presented in Figure \ref{fig:pynguin_smp}.}, label={fig:pynguin_smp}] 
def test_case_0():
    int_0 = -2973
    int_1 = 815
    bool_0 = module_0.any_int(int_0, int_0, int_1)
    assert bool_0 is False
\end{lstlisting}

\subsection{\textbf{Execution Time}}%\Amin{Execution or running time?}} 
The open-access API of Codex has a limit on the number of requests (20 per minute) and the number of tokens (40,000 per minute). For this reason, our experiment needs to stop calling the API once in a while to not exceed the limit.  As a result, we present the processing time analysis using llama-2-chat. The overall processing time of \tool{} on \textit{HumanEval} dataset while using llama-2-chat is on average 39.75 seconds with zero-shot learning (with a min of 16.16 and a max of 56.66 seconds) and 42.11 seconds with the few-shot prompt (with a min of 18.2 and a max of 64.2 seconds) per task. It includes on average building and calling initial prompts on LLMC with an average of 10.26 seconds, syntax fixing including calling the syntax fixing prompt on LLMC with 10.3 seconds, intended behavior repair at 0.38 seconds, MS calculation at 1.7 seconds, creating augmented prompts and calling them on LLM with 12.05 second and greedy optimization with 1.4 seconds. It is noteworthy that following the prompt augmentation step, \tool{} must reiterate the processes of syntax fixing, intended behavior repair, and greedy steps which are already included in the overall processing time. Among all steps of \tool{}, the most time-consuming ones are those that entail inferring output from the LLM. Conversely, the overall processing time on the same benchmark with Pynguin to complete searching the required space is on average 44.16 seconds with a min of 2.7 and a max of 10 minutes which is the default timeout of the tool.

\subsection{\textbf{The benefit of dialog LLM}} 
Our findings indicate that the dialog setup of llama-2-chat provides \tool{} with the flexibility to assign distinct roles to each component of the augmented prompt. For instance, by assigning \textit{IUT} to an assistant role during the prompt augmentation process, the likelihood of repeating the initial tests in the generated output is reduced, while the chance of generating new test cases for detecting surviving mutants is increased. Listing~\ref{fig:llama2-sample} illustrates an example of how llama-2-chat effectively synthesizes the difference of $PUT$ and one of its surviving mutants, explains the differences, and subsequently generates a new test case to detect the fault.

%However, Figure ~\ref{fig:survivedmut} on randomly choosing surviving mutants to augment the prompt  on different runs shows that the major gain in MS is achieved by augmenting the prompt with less than half of the mutants.

\vfill\null
%\columnbreak

\begin{lstlisting}[ language=python, style=mystyle, caption={A sample of augmented prompt and inference output on llama-2-chat.}, label={fig:llama2-sample}]
############
//prompt
############
{"role": "system", "content": "You are a Python 
coding assistant. Always answer with Python code."},
{"role": "user", "content":
#Generate test cases for the following code:
```python
def choose_num(x, y):
    if x > y:
        return -1
    if y % 2 == 0:
        return y
    if x == y:
        return -1
    return y - 1
```
},

{"role": "assistant", "content":
```python
#test cases
def test():
     assert choose_num(6,10) == 10
     assert choose_num(11,6) == -1
     assert choose_num(8,12) == 12
```
},

{"role": "user", "content":
#The test function, test(), cannot detect the fault in the following code:
```python
def choose_num(x, y):
    
    if x > y:
        return -1
    if y % 2 == 0:
        return y
    if x != y: 
        return -1
    return y - 1
```
#Provide a new test case to detect the fault in the prior code:
```python
#test case
def test():
    assert
}

##################################
//Generated output of llama-2-chat
##################################
"Here is a new test case to test the fault in the 
previous code:"
choose_num(5, 5) == -1

```python
def test(): 
     assertchoose_num(5, 5) == -1
```
"This test case checks for the scenario where both x 
and y are equal. In the modified code, the condition 
if x != y will evaluate to False, and the expected 
result is -1. However, in the original code,
the condition if x == y will evaluate to True, and 
the incorrect result would be y - 1, which is 4 
instead of -1."


\end{lstlisting}

\subsection{\textbf{Evalution Metrics}} 
Prior studies~\cite{tufano2022generating, nashidretrieval, dinella2022toga} that involve the generation of assertions through LLMs have employed the \textit{\enquote{exact match}} as one of their evaluation metrics. Exact match calculates the percentage of test cases generated by LLMs (the inferred output) that lexically match with the ground truth test cases (expected output). However, CIDAR~\cite{nashidretrieval} has already discussed the inadequacy of exact match as a suitable metric for assessing assertions produced by LLMs. This reason is that the model often generates assertions that are semantically correct %\Amin{do you mean good or proper assertions? if in the literature they use "accurate", then go for it Arghavn: in the paper they call it incorrect asssertions}
but may not precisely match the ground truth. In our study, \tool{} executed each test case including assertions, both on the $PUT$ and on its mutants to assess their correctness and effectiveness, reporting their MS. MS is a metric frequently used in prior studies and it serves as an effective metric for evaluating the quality of the test oracle~\cite{xie2006augmenting}. While, in this paper, we focus on improving the effectiveness of test cases in terms of fault detection, there are other metrics such as test coverage that can assess other quality aspects of a test case. Improving MS does not necessarily lead to good coverage and test coverage is weakly correlated with the efficiency of tests in fault detection \cite{CaiCovergae2005} and is challenged as a measure of test effectiveness in revealing faults \cite{gopinath2014code,hemmati2015effective}, which can make it challenging for our proposed method to perform well on both metrics~\cite{Papadakis19, fraser2010mutation}.

Furthermore, the results presented in~\cite{siddiq2023exploring} indicate that approximately 60\% of the test cases generated by Codex encounter compilation issues due to syntax errors. The incorporation of syntax correction and intended behavior repair steps in our proposed method, \tool{}, significantly enhances the utility of the tests generated by LLMs. 

\subsection{\textbf{Surviving mutants}} 
%\Amin{I only reorganized this part, no new part added, have a look:}
We augment the prompt at each iteration for each \textit{PUT} with a single surviving mutant. The average number of mutants for all \textit{PUT}s in \textit{HumanEval} and \textit{Refactory} are $6.6$ and $4.2$ and the average number of surviving mutants are $3.6$ and $1.8$, respectively. Using a combination of surviving mutants to augment the prompt could impact the speed of reaching $100\%$ MS. However, not all surviving mutants used in prompt augmentation contribute to improving MS, sometimes new test cases that address one mutant can also kill the remaining surviving mutants.%and as shown in Figure~\ref{fig:survivedmut}, the major gain in MS is achieved by augmenting the prompt with less than half of the mutants.

%%%%%%%%%%%%%%%%%%%%%%%%%%%
\section{Threats to Validity} \label{sec:thvalid}
\textbf{Internal validity.} %The MS computation may introduce some threats. While, in the literature, there are various mutation testing tools for Python, each comes with its own limitations and disadvantages as we did not find them suitable for our work. Therefore, we decided to compute the MS ourselves. 
%A threat to internal validity is due to potentially flaky tests; a test is called flaky when its verdict changes non-deterministically. It is acknowledged that flakiness could cause problems for Python test suites~\cite{gruber2021empirical} and automatically generated tests as well~\cite{fan2019systematic, parry2021survey}. However, our tests are not flaky based on our observations. \Foutse{what guarantee do you have that this will always be the case for all tests generated with Mutap? we need to cover for such criticism!} 

%Moreover, using closed-source LLM as a black box causes \Foutse{what does this sentence means? what do you mean by causing??? to improve??? are you trying to explain that it is unclear if model optimisations behind the api interfered with your obtained results? please formulate clear sentences!!!} adapting the prompt with different prompt-based learning techniques to improve the models' output. This may bias the outcome of the model as well \Foutse{please reformulate these sentences...they are poorly structured}. 

In this study, we employed two different prompt-based learning techniques: \textit{zero-shot} and \textit{few-shot}. However, we did not explore the potential impact of altering the natural language instructions or demonstrative examples (for \textit{few-shot} learning) within our prompts. Modifying these instructions or utilizing different demonstrative examples more closely aligned with the \textit{PUT}'s functionality could potentially enhance the results. As demonstrated by our results in RQ2, including the IUT in the prompt during augmentation steps reduced the instances of unintended behavior in test oracles.  Conversely, using, for example, lengthy natural language instructions might potentially have an adverse effect on the results. 
%\Foutse{so? how does this impact your reported results? would the reported efficiency of Mutap still holds????}

To repair syntax errors in test cases through re-prompting the LLMC, we have employed the approach presented in~\cite{zhang2022repairing}. We did not integrate additional information about the syntax error such as error messages or error lines into the prompt. It is worth considering that incorporating additional information about syntax errors could potentially enhance the LLMC's performance to repair these syntax errors.

%About fixing syntax errors in test cases by re-prompting the LLMC, we have used the strategy proposed in~\cite{zhang2022repairing}, without benefiting from the error messages. while including extra information about syntax errors such as the details of error messages may improve the capability of the LLMC to fix the syntax error.%\Foutse{what does that means? does it means that it is more or less effective??}. Although the focus of our study is not on fixing errors, having an error message in the prompt of the syntax fixer may improve the capability of the model to fix the syntax error. 

Additionally, we acknowledge that the greedy algorithm employed in our approach to minimize the number of test oracles might not be the most optimal solution for minimizing test oracles while maximizing MS. However, prior studies~\cite{fraser2010mutation,fraser2011evosuite} using the same method to minimize the number of assertions have demonstrated its effectiveness in reducing the number of test oracles within test cases, along with its ease of implementation.  
%\Foutse{which greedy approach? used where? be precise please!!!!}
%\Foutse{...etc...suggesting that....etc please complete adequately!} %that have used this technique to minimize the number of assertions.

Finally, among different types of assertions, we only focus on generating primitive ones in this study. Other assertion types can be explored in future studies.

%\Amin{was it the case for us: To mitigate this threat, generating mutants is done using a customized version of the MUTPY mutation testing tool.}
We employ the notions of mutant killability and bug detection as metrics to gauge the effectiveness of test cases, given that the primary objective of testing is to uncover bugs. Coverage has been employed in various other studies to assess test case quality~\cite{schafer2023adaptive,lemieux2023codamosa}. However, it has been demonstrated that while there exists a correlation between coverage and bug detection, they may not consistently align in ranking different testing strategies, as observed in the realm of fuzz testing~\cite{bohme2022reliability}.

\textbf{Construct Validity.} We use the notions of killing mutants and bug detection as metrics to evaluate the effectiveness of test cases, given that the primary objective of testing is to reveal bugs. Coverage has been employed in various other studies to assess test case quality~\cite{schafer2023adaptive,lemieux2023codamosa}. It has been shown that although there is a correlation between coverage and bug-finding, they do not agree on the ranking of different testers, like in fuzz testing space~\cite{bohme2022reliability}. %Also, SBST demonstrated \Foutse{demonstration???} the difficulty in finding bugs, especially where the construction of complex objects and generation of specific primitives are required \cite{almasi2017industrial}. 

It's important to note that the bugs present in mutants are artificial and might not directly correspond to real-world faults. To address this concern, we've employed the \textit{Refactory}~\cite{hu2019re} dataset, a bug-repairing benchmark that contains real faulty programs developed by students.

\textbf{External Validity.} For our experiments, we used two datasets containing Python programming tasks, which could potentially pose external challenges to the validity of our findings. The requirement for executable Python programs is essential to run the generated tests against both the accurate and buggy versions (real or mutated) of \textit{PUT} and this consideration guided our choice of datasets.
However, since we didn't make any specific assumptions while selecting the dataset, our results can be extended to other Python programs. %\Amin{Arghavan, please check this:}Since some of the generated mutants were not runnable, they lead to hang or crash. Therefore, we were not able to assess the assertions against those mutants and excluded them from our experiments and calculation of MS. 

Finally, it should be acknowledged that the technique proposed and the evaluations conducted in this paper are conceptually adaptable to languages beyond Python. However, the current implementation of \tool{} is tailored for Python programs, meaning our existing results cannot be extended to cover other programming languages.

\textbf{Reliability validity.} For the purpose of enabling other researchers to replicate or expand upon our study, we provide a replication package \cite{rep-package}. However, the ongoing enhancement of LLMs could potentially pose a challenge to achieving an exact replication of our results.

%%%%%%%%%%%%%%%%%%%%%%%%%%%%%%%%%
\section{Related work} \label{sec:related}
Authors in~\cite{bareiss2022code} studied the impact of \textit{few-shot} learning across various downstream tasks, including test case and test oracle generation. They compared the performance of \textit{few-shot} learning with automatic test generation tools. The investigation was conducted on a different set of Java methods sourced from different benchmarks.  
The outcomes indicated that LLMs possess the capability to generate test cases and test oracles that exactly match (in lexical terms) the ground truth tests within the benchmark projects. Furthermore, their test coverage was found to be comparable with test cases generated by automatic test generation tools.

Sch{"a}fer et al.~\cite{schafer2023adaptive} undertook an effort to generate test cases by prompting Codex. Their investigation was concentrated on $25$ JavaScript packages. The prompt in their study encompassed the implementation of the PUT and also the usage examples of APIs extracted from documentation. In instances where a test case proved unsuccessful on the PUT, their method incorporated the encountered error message into the prompt and re-prompted Codex. Their findings demonstrated that the process of enhancing the prompt with such additional information facilitated Codex in producing correct test cases with sufficient coverage.

LIBRO~\cite{kang2022large} used the issue reports (both title and body) as \textit{few-shot} prompts to generate bug-reproducing test cases. The final test cases were incorporated into appropriate test classes and ranked based on their validity. The results revealed an enhancement in generating correct test cases to reproduce bugs compared to state-of-the-art tools.

CEDAR~\cite{nashidretrieval}, rather than employing fixed demonstrative examples in \textit{few-shot} learning, aimed to retrieve demonstrative examples related to each \textit{PUT} and incorporate them into the prompt. They assessed their method based on the lexical match,  termed "exact match", between generated assertions and the ground truth in a benchmark. While their proposed approach demonstrates enhanced performance in achieving exact matches between assertions and the ground truth, it necessitates an extensive pull of code samples for the selection of appropriate demonstrative examples for each \textit{PUT}.

ATHENATEST~\cite{tufano2021unit} employed the BART transformer model~\cite{lewis2019bart}, which they fine-tuned using a collection of Java functions and their corresponding tests. They reported test coverage comparable to those of EvoSuite~\cite{fraser2011evosuite} upon evaluating generating test cases for five Java projects.

TOGA~\cite{dinella2022toga} engaged in fine-tuning CodeBERT using the \textit{PUT}'s docstring along with the prefix of a test case featuring a masked assertion. Their goal was to synthesize the assertion. Subsequently, they formulated the whole test oracles by incorporating a test oracle grammar and generating a set of assertions. This set was then subjected to ranking through a neural network ranker based on their lexical match to ground truth test oracles. Although they reported results akin to those of EvoSuite~\cite{fraser2011evosuite} in bug detection, their focus is only on synthesizing the assertions.
However, synthesizing assertion is not challenging but generating effective and meaningful test oracles poses a significant challenge. 

CODAMOSA combined the test cases generated by Codex with those derived from Pynguin in cases where Pynguin's test case generation halted and failed to enhance test coverage.  CODAMOSA achieves higher test coverage on various Python benchmarks~\cite{lemieux2023codamosa} compared to Pynguin. It is worth noting that, akin to other studies, CODAMOSA concentrated solely on test coverage improvement, and its generated test cases lacked assertion oracles for bug detection within programs.

Two additional studies employed Codex to simultaneously generate code and corresponding test cases based on a given problem description. Subsequently, they used these test cases to filter out buggy suggestions produced by Codex~\cite{lahiri2022interactive, chen2022codet}. For code generation, they employed the problem description as a prompt, and for test case generation, they used the same problem description along with the \textit{PUT} and a natural language instruction.

Although prior research has explored diverse strategies for generating test cases using LLMs like Codex and assessed them in terms of test coverage or lexical match with ground truth tests, none of these studies specifically focused on leveraging MT to enhance the effectiveness of the generated test cases.

%%%%%%%%%%%%%%%%%%%%%%%%%%%%%%%%%%%
\section{Conclusion} \label{sec:conclusion}
In this paper, we proposed \tool{} as a means of improving and evaluating the ability of pre-trained LLMs to generate effective test cases. \tool{} first prompts its LLMC to generate test cases using \textit{zero-shot} and \textit{few-shot} learning. After identifying and correcting any potential syntax and return value errors in the generated test cases, \tool{} evaluates their effectiveness by conducting MT. Then, it uses the surviving mutants of each \textit{PUT}, if any, as well as the initial inadequate test case to augment the initial prompt. It re-prompts its LLMC using the augmented prompt to regenerate new test cases that are capable of detecting surviving mutants. 

We assessed the effectiveness of the test cases generated by LLMs to identify bugs in real and synthetic buggy programs. On average, test cases generated by \tool{} successfully identify $86.72\%$ of buggy code in a bug repairing benchmark when using the LLM designed for code generation, Codex. When employing LLM with the dialog setup, llama-2-chat, \tool{} further improves its performance, detecting $94.06\%$ of the buggy code, outperforming both an automatic test generation tool and a \textit{zero-shot} and \textit{few-shot} learning technique on LLMs. This underscores the advantage of employing LLMs as the core of an automatic test generation tool, as conventional automatic generation tools such as Pynguin lack access to the insights embedded in surviving mutants.

Although the current version of \tool{} employs two different LLMs to generate test cases for Python programs, its design and evaluation methodology are fundamentally adaptable to various programming languages and models. Therefore, as future work, it can be easily expanded to encompass other programming languages or incorporate new LLMs.

%%%%%%%%%%%%%%%%%%%%%%%%%%%%%%%%%%%%%%%%%%%%%%%
%$%%%%%%%%%%%%%%%%%%%%%%%%%%%%

\printcredits

\bibliographystyle{cas-model2-names}
\balance
\bibliography{cas-refs}

\begin{thebibliography}{52}
\expandafter\ifx\csname natexlab\endcsname\relax\def\natexlab#1{#1}\fi
\providecommand{\url}[1]{\texttt{#1}}
\providecommand{\href}[2]{#2}
\providecommand{\path}[1]{#1}
\providecommand{\DOIprefix}{doi:}
\providecommand{\ArXivprefix}{arXiv:}
\providecommand{\URLprefix}{URL: }
\providecommand{\Pubmedprefix}{pmid:}
\providecommand{\doi}[1]{\href{http://dx.doi.org/#1}{\path{#1}}}
\providecommand{\Pubmed}[1]{\href{pmid:#1}{\path{#1}}}
\providecommand{\bibinfo}[2]{#2}
\ifx\xfnm\relax \def\xfnm[#1]{\unskip,\space#1}\fi
%Type = Article
\bibitem[{Ahmed and Devanbu(2022)}]{ahmed2022few}
\bibinfo{author}{Ahmed, T.}, \bibinfo{author}{Devanbu, P.},
  \bibinfo{year}{2022}.
\newblock \bibinfo{title}{Few-shot training {LLM}s for project-specific
  code-summarization}.
\newblock \bibinfo{journal}{arXiv preprint arXiv:2207.04237} .
%Type = Misc
\bibitem[{Alvarado et~al.(2023)Alvarado, Gazit and Wattenberger}]{copilotlab}
\bibinfo{author}{Alvarado, I.}, \bibinfo{author}{Gazit, I.},
  \bibinfo{author}{Wattenberger, A.}, \bibinfo{year}{2023}.
\newblock \bibinfo{title}{Github copilot labs}.
\newblock
  \bibinfo{howpublished}{\url{https://githubnext.com/projects/copilot-labs/}}.
%Type = Misc
\bibitem[{Anonymous(2023)}]{rep-package}
\bibinfo{author}{Anonymous}, \bibinfo{year}{2023}.
\newblock \bibinfo{title}{The replication package}.
\newblock
  \bibinfo{howpublished}{\url{https://github.com/ExpertiseModel/MuTAP}}.
%Type = Article
\bibitem[{Arcuri(2018)}]{arcuri2018test}
\bibinfo{author}{Arcuri, A.}, \bibinfo{year}{2018}.
\newblock \bibinfo{title}{Test suite generation with the {M}any {I}ndependent
  {O}bjective ({MIO}) algorithm}.
\newblock \bibinfo{journal}{Information and Software Technology}
  \bibinfo{volume}{104}, \bibinfo{pages}{195--206}.
%Type = Article
\bibitem[{Arcuri and Fraser(2013)}]{arcuri2013parameter}
\bibinfo{author}{Arcuri, A.}, \bibinfo{author}{Fraser, G.},
  \bibinfo{year}{2013}.
\newblock \bibinfo{title}{Parameter tuning or default values? an empirical
  investigation in search-based software engineering}.
\newblock \bibinfo{journal}{Empirical Software Engineering}
  \bibinfo{volume}{18}, \bibinfo{pages}{594--623}.
%Type = Inproceedings
\bibitem[{Arteca et~al.(2022)Arteca, Harner, Pradel and Tip}]{arteca2022nessie}
\bibinfo{author}{Arteca, E.}, \bibinfo{author}{Harner, S.},
  \bibinfo{author}{Pradel, M.}, \bibinfo{author}{Tip, F.},
  \bibinfo{year}{2022}.
\newblock \bibinfo{title}{Nessie: automatically testing javascript apis with
  asynchronous callbacks}, in: \bibinfo{booktitle}{Proceedings of the 44th
  International Conference on Software Engineering}, pp.
  \bibinfo{pages}{1494--1505}.
%Type = Article
\bibitem[{Barei{\ss} et~al.(2022)Barei{\ss}, Souza, d'Amorim and
  Pradel}]{bareiss2022code}
\bibinfo{author}{Barei{\ss}, P.}, \bibinfo{author}{Souza, B.},
  \bibinfo{author}{d'Amorim, M.}, \bibinfo{author}{Pradel, M.},
  \bibinfo{year}{2022}.
\newblock \bibinfo{title}{Code generation tools (almost) for free? a study of
  few-shot, pre-trained language models on code}.
\newblock \bibinfo{journal}{arXiv preprint arXiv:2206.01335} .
%Type = Inproceedings
\bibitem[{B{\"o}hme et~al.(2022)B{\"o}hme, Szekeres and
  Metzman}]{bohme2022reliability}
\bibinfo{author}{B{\"o}hme, M.}, \bibinfo{author}{Szekeres, L.},
  \bibinfo{author}{Metzman, J.}, \bibinfo{year}{2022}.
\newblock \bibinfo{title}{On the reliability of coverage-based fuzzer
  benchmarking}, in: \bibinfo{booktitle}{Proceedings of the 44th International
  Conference on Software Engineering}, pp. \bibinfo{pages}{1621--1633}.
%Type = Article
\bibitem[{Brown et~al.(2020)Brown, Mann, Ryder, Subbiah, Kaplan, Dhariwal,
  Neelakantan, Shyam, Sastry, Askell et~al.}]{brown2020language}
\bibinfo{author}{Brown, T.}, \bibinfo{author}{Mann, B.},
  \bibinfo{author}{Ryder, N.}, \bibinfo{author}{Subbiah, M.},
  \bibinfo{author}{Kaplan, J.D.}, \bibinfo{author}{Dhariwal, P.},
  \bibinfo{author}{Neelakantan, A.}, \bibinfo{author}{Shyam, P.},
  \bibinfo{author}{Sastry, G.}, \bibinfo{author}{Askell, A.}, et~al.,
  \bibinfo{year}{2020}.
\newblock \bibinfo{title}{Language models are few-shot learners}.
\newblock \bibinfo{journal}{Advances in neural information processing systems}
  \bibinfo{volume}{33}, \bibinfo{pages}{1877--1901}.
%Type = Inproceedings
\bibitem[{Cai and Lyu(2005)}]{CaiCovergae2005}
\bibinfo{author}{Cai, X.}, \bibinfo{author}{Lyu, M.R.}, \bibinfo{year}{2005}.
\newblock \bibinfo{title}{The effect of code coverage on fault detection under
  different testing profiles}, in: \bibinfo{booktitle}{Proceedings of the 1st
  International Workshop on Advances in Model-Based Testing},
  \bibinfo{publisher}{ACM}, \bibinfo{address}{New York, NY, USA}. p.
  \bibinfo{pages}{1–7}.
\newblock \URLprefix \url{https://doi.org/10.1145/1083274.1083288}.
%Type = Article
\bibitem[{Chen et~al.(2022)Chen, Zhang, Nguyen, Zan, Lin, Lou and
  Chen}]{chen2022codet}
\bibinfo{author}{Chen, B.}, \bibinfo{author}{Zhang, F.},
  \bibinfo{author}{Nguyen, A.}, \bibinfo{author}{Zan, D.},
  \bibinfo{author}{Lin, Z.}, \bibinfo{author}{Lou, J.G.},
  \bibinfo{author}{Chen, W.}, \bibinfo{year}{2022}.
\newblock \bibinfo{title}{Codet: Code generation with generated tests}.
\newblock \bibinfo{journal}{arXiv preprint arXiv:2207.10397} .
%Type = Article
\bibitem[{Chen et~al.(2021)Chen, Tworek, Jun, Yuan, Pinto, Kaplan, Edwards,
  Burda, Joseph, Brockman et~al.}]{chen2021evaluating}
\bibinfo{author}{Chen, M.}, \bibinfo{author}{Tworek, J.}, \bibinfo{author}{Jun,
  H.}, \bibinfo{author}{Yuan, Q.}, \bibinfo{author}{Pinto, H.P.d.O.},
  \bibinfo{author}{Kaplan, J.}, \bibinfo{author}{Edwards, H.},
  \bibinfo{author}{Burda, Y.}, \bibinfo{author}{Joseph, N.},
  \bibinfo{author}{Brockman, G.}, et~al., \bibinfo{year}{2021}.
\newblock \bibinfo{title}{Evaluating large language models trained on code}.
\newblock \bibinfo{journal}{arXiv preprint arXiv:2107.03374} .
%Type = Article
\bibitem[{Clement et~al.(2020)Clement, Drain, Timcheck, Svyatkovskiy and
  Sundaresan}]{clement2020pymt5}
\bibinfo{author}{Clement, C.B.}, \bibinfo{author}{Drain, D.},
  \bibinfo{author}{Timcheck, J.}, \bibinfo{author}{Svyatkovskiy, A.},
  \bibinfo{author}{Sundaresan, N.}, \bibinfo{year}{2020}.
\newblock \bibinfo{title}{Pymt5: multi-mode translation of natural language and
  python code with transformers}.
\newblock \bibinfo{journal}{arXiv preprint arXiv:2010.03150} .
%Type = Inproceedings
\bibitem[{Dinella et~al.(2022)Dinella, Ryan, Mytkowicz and
  Lahiri}]{dinella2022toga}
\bibinfo{author}{Dinella, E.}, \bibinfo{author}{Ryan, G.},
  \bibinfo{author}{Mytkowicz, T.}, \bibinfo{author}{Lahiri, S.K.},
  \bibinfo{year}{2022}.
\newblock \bibinfo{title}{Toga: a neural method for test oracle generation},
  in: \bibinfo{booktitle}{Proceedings of the 44th International Conference on
  Software Engineering}, pp. \bibinfo{pages}{2130--2141}.
%Type = Article
\bibitem[{Feng et~al.(2020)Feng, Guo, Tang, Duan, Feng, Gong, Shou, Qin, Liu,
  Jiang et~al.}]{feng2020codebert}
\bibinfo{author}{Feng, Z.}, \bibinfo{author}{Guo, D.}, \bibinfo{author}{Tang,
  D.}, \bibinfo{author}{Duan, N.}, \bibinfo{author}{Feng, X.},
  \bibinfo{author}{Gong, M.}, \bibinfo{author}{Shou, L.}, \bibinfo{author}{Qin,
  B.}, \bibinfo{author}{Liu, T.}, \bibinfo{author}{Jiang, D.}, et~al.,
  \bibinfo{year}{2020}.
\newblock \bibinfo{title}{Codebert: A pre-trained model for programming and
  natural languages}.
\newblock \bibinfo{journal}{arXiv preprint arXiv:2002.08155} .
%Type = Inproceedings
\bibitem[{Fraser and Arcuri(2011a)}]{fraser2011evolutionary}
\bibinfo{author}{Fraser, G.}, \bibinfo{author}{Arcuri, A.},
  \bibinfo{year}{2011}a.
\newblock \bibinfo{title}{Evolutionary generation of whole test suites}, in:
  \bibinfo{booktitle}{2011 11th International Conference on Quality Software},
  \bibinfo{organization}{IEEE}. pp. \bibinfo{pages}{31--40}.
%Type = Inproceedings
\bibitem[{Fraser and Arcuri(2011b)}]{fraser2011evosuite}
\bibinfo{author}{Fraser, G.}, \bibinfo{author}{Arcuri, A.},
  \bibinfo{year}{2011}b.
\newblock \bibinfo{title}{Evosuite: automatic test suite generation for
  object-oriented software}, in: \bibinfo{booktitle}{Proceedings of the 19th
  ACM SIGSOFT symposium and the 13th European conference on Foundations of
  software engineering}, pp. \bibinfo{pages}{416--419}.
%Type = Inproceedings
\bibitem[{Fraser and Zeller(2010)}]{fraser2010mutation}
\bibinfo{author}{Fraser, G.}, \bibinfo{author}{Zeller, A.},
  \bibinfo{year}{2010}.
\newblock \bibinfo{title}{Mutation-driven generation of unit tests and
  oracles}, in: \bibinfo{booktitle}{Proceedings of the 19th international
  symposium on Software testing and analysis}, pp. \bibinfo{pages}{147--158}.
%Type = Inproceedings
\bibitem[{Godefroid et~al.(2005)Godefroid, Klarlund and
  Sen}]{godefroid2005dart}
\bibinfo{author}{Godefroid, P.}, \bibinfo{author}{Klarlund, N.},
  \bibinfo{author}{Sen, K.}, \bibinfo{year}{2005}.
\newblock \bibinfo{title}{Dart: Directed automated random testing}, in:
  \bibinfo{booktitle}{Proceedings of the 2005 ACM SIGPLAN conference on
  Programming language design and implementation}, pp.
  \bibinfo{pages}{213--223}.
%Type = Inproceedings
\bibitem[{Gopinath et~al.(2014)Gopinath, Jensen and Groce}]{gopinath2014code}
\bibinfo{author}{Gopinath, R.}, \bibinfo{author}{Jensen, C.},
  \bibinfo{author}{Groce, A.}, \bibinfo{year}{2014}.
\newblock \bibinfo{title}{Code coverage for suite evaluation by developers},
  in: \bibinfo{booktitle}{Proceedings of the 36th International Conference on
  Software Engineering}, pp. \bibinfo{pages}{72--82}.
%Type = Misc
\bibitem[{Ha{\l}as(2019)}]{mutpy}
\bibinfo{author}{Ha{\l}as, K.}, \bibinfo{year}{2019}.
\newblock \bibinfo{title}{Mutpy: a mutation testing tool for {P}ython 3.x
  source code}.
\newblock \bibinfo{howpublished}{\url{https://github.com/mutpy/mutpy}}.
%Type = Inproceedings
\bibitem[{Hemmati(2015)}]{hemmati2015effective}
\bibinfo{author}{Hemmati, H.}, \bibinfo{year}{2015}.
\newblock \bibinfo{title}{How effective are code coverage criteria?}, in:
  \bibinfo{booktitle}{2015 IEEE International Conference on Software Quality,
  Reliability and Security}, \bibinfo{organization}{IEEE}. pp.
  \bibinfo{pages}{151--156}.
%Type = Inproceedings
\bibitem[{Hu et~al.(2019)Hu, Ahmed, Mechtaev, Leong and
  Roychoudhury}]{hu2019re}
\bibinfo{author}{Hu, Y.}, \bibinfo{author}{Ahmed, U.Z.},
  \bibinfo{author}{Mechtaev, S.}, \bibinfo{author}{Leong, B.},
  \bibinfo{author}{Roychoudhury, A.}, \bibinfo{year}{2019}.
\newblock \bibinfo{title}{Re-factoring based program repair applied to
  programming assignments}, in: \bibinfo{booktitle}{2019 34th IEEE/ACM
  International Conference on Automated Software Engineering (ASE)},
  \bibinfo{organization}{IEEE}. pp. \bibinfo{pages}{388--398}.
%Type = Misc
\bibitem[{Hu et~al.(2023)Hu, Ahmed, Mechtaev, Leong and
  Roychoudhury}]{re-bug-repair}
\bibinfo{author}{Hu, Y.}, \bibinfo{author}{Ahmed, U.Z.},
  \bibinfo{author}{Mechtaev, S.}, \bibinfo{author}{Leong, B.},
  \bibinfo{author}{Roychoudhury, A.}, \bibinfo{year}{2023}.
\newblock \bibinfo{title}{Refactory}.
\newblock
  \bibinfo{howpublished}{\url{https://github.com/githubhuyang/refactory}}.
%Type = Article
\bibitem[{Jia and Harman(2011)}]{Jia11}
\bibinfo{author}{Jia, Y.}, \bibinfo{author}{Harman, M.}, \bibinfo{year}{2011}.
\newblock \bibinfo{title}{An analysis and survey of the development of mutation
  testing}.
\newblock \bibinfo{journal}{IEEE Transactions on Software Engineering}
  \bibinfo{volume}{37}, \bibinfo{pages}{649--678}.
\newblock \DOIprefix\doi{10.1109/TSE.2010.62}.
%Type = Article
\bibitem[{Joshi et~al.(2022)Joshi, Cambronero, Gulwani, Le, Radicek and
  Verbruggen}]{joshi2022repair}
\bibinfo{author}{Joshi, H.}, \bibinfo{author}{Cambronero, J.},
  \bibinfo{author}{Gulwani, S.}, \bibinfo{author}{Le, V.},
  \bibinfo{author}{Radicek, I.}, \bibinfo{author}{Verbruggen, G.},
  \bibinfo{year}{2022}.
\newblock \bibinfo{title}{Repair is nearly generation: Multilingual program
  repair with {LLM}s}.
\newblock \bibinfo{journal}{arXiv preprint arXiv:2208.11640} .
%Type = Article
\bibitem[{Kang et~al.(2022)Kang, Yoon and Yoo}]{kang2022large}
\bibinfo{author}{Kang, S.}, \bibinfo{author}{Yoon, J.}, \bibinfo{author}{Yoo,
  S.}, \bibinfo{year}{2022}.
\newblock \bibinfo{title}{Large language models are few-shot testers: Exploring
  {LLM}-based general bug reproduction}.
\newblock \bibinfo{journal}{arXiv preprint arXiv:2209.11515} .
%Type = Article
\bibitem[{Lahiri et~al.(2022)Lahiri, Naik, Sakkas, Choudhury, von Veh,
  Musuvathi, Inala, Wang and Gao}]{lahiri2022interactive}
\bibinfo{author}{Lahiri, S.K.}, \bibinfo{author}{Naik, A.},
  \bibinfo{author}{Sakkas, G.}, \bibinfo{author}{Choudhury, P.},
  \bibinfo{author}{von Veh, C.}, \bibinfo{author}{Musuvathi, M.},
  \bibinfo{author}{Inala, J.P.}, \bibinfo{author}{Wang, C.},
  \bibinfo{author}{Gao, J.}, \bibinfo{year}{2022}.
\newblock \bibinfo{title}{Interactive code generation via test-driven
  user-intent formalization}.
\newblock \bibinfo{journal}{arXiv preprint arXiv:2208.05950} .
%Type = Inproceedings
\bibitem[{Lemieux et~al.(2023)Lemieux, Inala, Lahiri and
  Sen}]{lemieux2023codamosa}
\bibinfo{author}{Lemieux, C.}, \bibinfo{author}{Inala, J.P.},
  \bibinfo{author}{Lahiri, S.K.}, \bibinfo{author}{Sen, S.},
  \bibinfo{year}{2023}.
\newblock \bibinfo{title}{Codamosa: Escaping coverage plateaus in test
  generation with pre-trained large language models}, in:
  \bibinfo{booktitle}{Accepted by 45th International Conference on Software
  Engineering (ICSE)}.
%Type = Article
\bibitem[{Lewis et~al.(2019)Lewis, Liu, Goyal, Ghazvininejad, Mohamed, Levy,
  Stoyanov and Zettlemoyer}]{lewis2019bart}
\bibinfo{author}{Lewis, M.}, \bibinfo{author}{Liu, Y.}, \bibinfo{author}{Goyal,
  N.}, \bibinfo{author}{Ghazvininejad, M.}, \bibinfo{author}{Mohamed, A.},
  \bibinfo{author}{Levy, O.}, \bibinfo{author}{Stoyanov, V.},
  \bibinfo{author}{Zettlemoyer, L.}, \bibinfo{year}{2019}.
\newblock \bibinfo{title}{Bart: Denoising sequence-to-sequence pre-training for
  natural language generation, translation, and comprehension}.
\newblock \bibinfo{journal}{arXiv preprint arXiv:1910.13461} .
%Type = Article
\bibitem[{Liu et~al.(2023)Liu, Yuan, Fu, Jiang, Hayashi and
  Neubig}]{liu2023pre}
\bibinfo{author}{Liu, P.}, \bibinfo{author}{Yuan, W.}, \bibinfo{author}{Fu,
  J.}, \bibinfo{author}{Jiang, Z.}, \bibinfo{author}{Hayashi, H.},
  \bibinfo{author}{Neubig, G.}, \bibinfo{year}{2023}.
\newblock \bibinfo{title}{Pre-train, prompt, and predict: A systematic survey
  of prompting methods in natural language processing}.
\newblock \bibinfo{journal}{ACM Computing Surveys} \bibinfo{volume}{55},
  \bibinfo{pages}{1--35}.
%Type = Inproceedings
\bibitem[{Lukasczyk and Fraser(2022)}]{lukasczyk2022pynguin}
\bibinfo{author}{Lukasczyk, S.}, \bibinfo{author}{Fraser, G.},
  \bibinfo{year}{2022}.
\newblock \bibinfo{title}{Pynguin: Automated unit test generation for python},
  in: \bibinfo{booktitle}{Proceedings of the ACM/IEEE 44th International
  Conference on Software Engineering: Companion Proceedings}, pp.
  \bibinfo{pages}{168--172}.
%Type = Article
\bibitem[{Lukasczyk et~al.(2023)Lukasczyk, Kroi{\ss} and
  Fraser}]{lukasczyk2023empirical}
\bibinfo{author}{Lukasczyk, S.}, \bibinfo{author}{Kroi{\ss}, F.},
  \bibinfo{author}{Fraser, G.}, \bibinfo{year}{2023}.
\newblock \bibinfo{title}{An empirical study of automated unit test generation
  for python}.
\newblock \bibinfo{journal}{Empirical Software Engineering}
  \bibinfo{volume}{28}, \bibinfo{pages}{36}.
%Type = Article
\bibitem[{{Moradi Dakhel} et~al.(2023){Moradi Dakhel}, Majdinasab, Nikanjam,
  Khomh, Desmarais and Jiang}]{MoradiJSS2023}
\bibinfo{author}{{Moradi Dakhel}, A.}, \bibinfo{author}{Majdinasab, V.},
  \bibinfo{author}{Nikanjam, A.}, \bibinfo{author}{Khomh, F.},
  \bibinfo{author}{Desmarais, M.C.}, \bibinfo{author}{Jiang, Z.M.J.},
  \bibinfo{year}{2023}.
\newblock \bibinfo{title}{Github {C}opilot {AI} pair programmer: Asset or
  liability?}
\newblock \bibinfo{journal}{Journal of Systems and Software}
  \bibinfo{volume}{203}, \bibinfo{pages}{111734}.
\newblock \DOIprefix\doi{https://doi.org/10.1016/j.jss.2023.111734}.
%Type = Article
\bibitem[{Nashid et~al.(2023)Nashid, Sintaha and Mesbah}]{nashidretrieval}
\bibinfo{author}{Nashid, N.}, \bibinfo{author}{Sintaha, M.},
  \bibinfo{author}{Mesbah, A.}, \bibinfo{year}{2023}.
\newblock \bibinfo{title}{Retrieval-based prompt selection for code-related
  few-shot learning} .
%Type = Inproceedings
\bibitem[{Palomba et~al.(2016)Palomba, Di~Nucci, Panichella, Oliveto and
  De~Lucia}]{palomba2016diffusion}
\bibinfo{author}{Palomba, F.}, \bibinfo{author}{Di~Nucci, D.},
  \bibinfo{author}{Panichella, A.}, \bibinfo{author}{Oliveto, R.},
  \bibinfo{author}{De~Lucia, A.}, \bibinfo{year}{2016}.
\newblock \bibinfo{title}{On the diffusion of test smells in automatically
  generated test code: An empirical study}, in: \bibinfo{booktitle}{Proceedings
  of the 9th international workshop on search-based software testing}, pp.
  \bibinfo{pages}{5--14}.
%Type = Inproceedings
\bibitem[{Panichella et~al.(2015)Panichella, Kifetew and
  Tonella}]{panichella2015reformulating}
\bibinfo{author}{Panichella, A.}, \bibinfo{author}{Kifetew, F.M.},
  \bibinfo{author}{Tonella, P.}, \bibinfo{year}{2015}.
\newblock \bibinfo{title}{Reformulating branch coverage as a many-objective
  optimization problem}, in: \bibinfo{booktitle}{2015 IEEE 8th international
  conference on software testing, verification and validation (ICST)},
  \bibinfo{organization}{IEEE}. pp. \bibinfo{pages}{1--10}.
%Type = Article
\bibitem[{Panichella et~al.(2017)Panichella, Kifetew and
  Tonella}]{panichella2017automated}
\bibinfo{author}{Panichella, A.}, \bibinfo{author}{Kifetew, F.M.},
  \bibinfo{author}{Tonella, P.}, \bibinfo{year}{2017}.
\newblock \bibinfo{title}{Automated test case generation as a many-objective
  optimisation problem with dynamic selection of the targets}.
\newblock \bibinfo{journal}{IEEE Transactions on Software Engineering}
  \bibinfo{volume}{44}, \bibinfo{pages}{122--158}.
%Type = Inproceedings
\bibitem[{Panichella et~al.(2020)Panichella, Panichella, Fraser, Sawant and
  Hellendoorn}]{panichella2020revisiting}
\bibinfo{author}{Panichella, A.}, \bibinfo{author}{Panichella, S.},
  \bibinfo{author}{Fraser, G.}, \bibinfo{author}{Sawant, A.A.},
  \bibinfo{author}{Hellendoorn, V.J.}, \bibinfo{year}{2020}.
\newblock \bibinfo{title}{Revisiting test smells in automatically generated
  tests: limitations, pitfalls, and opportunities}, in:
  \bibinfo{booktitle}{2020 IEEE international conference on software
  maintenance and evolution (ICSME)}, \bibinfo{organization}{IEEE}. pp.
  \bibinfo{pages}{523--533}.
%Type = Incollection
\bibitem[{Papadakis et~al.(2019)Papadakis, Kintis, Zhang, Jia, Le~Traon and
  Harman}]{Papadakis19}
\bibinfo{author}{Papadakis, M.}, \bibinfo{author}{Kintis, M.},
  \bibinfo{author}{Zhang, J.}, \bibinfo{author}{Jia, Y.},
  \bibinfo{author}{Le~Traon, Y.}, \bibinfo{author}{Harman, M.},
  \bibinfo{year}{2019}.
\newblock \bibinfo{title}{Mutation testing advances: an analysis and survey},
  in: \bibinfo{booktitle}{Advances in Computers}.
  \bibinfo{publisher}{Elsevier}. volume \bibinfo{volume}{112}, pp.
  \bibinfo{pages}{275--378}.
%Type = Article
\bibitem[{Sch{\"a}fer et~al.(2023)Sch{\"a}fer, Nadi, Eghbali and
  Tip}]{schafer2023adaptive}
\bibinfo{author}{Sch{\"a}fer, M.}, \bibinfo{author}{Nadi, S.},
  \bibinfo{author}{Eghbali, A.}, \bibinfo{author}{Tip, F.},
  \bibinfo{year}{2023}.
\newblock \bibinfo{title}{Adaptive test generation using a large language
  model}.
\newblock \bibinfo{journal}{arXiv preprint arXiv:2302.06527} .
%Type = Article
\bibitem[{Selakovic et~al.(2018)Selakovic, Pradel, Karim and
  Tip}]{selakovic2018test}
\bibinfo{author}{Selakovic, M.}, \bibinfo{author}{Pradel, M.},
  \bibinfo{author}{Karim, R.}, \bibinfo{author}{Tip, F.}, \bibinfo{year}{2018}.
\newblock \bibinfo{title}{Test generation for higher-order functions in dynamic
  languages}.
\newblock \bibinfo{journal}{Proceedings of the ACM on Programming Languages}
  \bibinfo{volume}{2}, \bibinfo{pages}{1--27}.
%Type = Article
\bibitem[{Sen et~al.(2005)Sen, Marinov and Agha}]{sen2005cute}
\bibinfo{author}{Sen, K.}, \bibinfo{author}{Marinov, D.},
  \bibinfo{author}{Agha, G.}, \bibinfo{year}{2005}.
\newblock \bibinfo{title}{Cute: A concolic unit testing engine for c}.
\newblock \bibinfo{journal}{ACM SIGSOFT Software Engineering Notes}
  \bibinfo{volume}{30}, \bibinfo{pages}{263--272}.
%Type = Book
\bibitem[{Shore and Warden(2021)}]{shore2021art}
\bibinfo{author}{Shore, J.}, \bibinfo{author}{Warden, S.},
  \bibinfo{year}{2021}.
\newblock \bibinfo{title}{The art of agile development}.
\newblock \bibinfo{edition}{2nd} ed., \bibinfo{publisher}{"O'Reilly"}.
%Type = Article
\bibitem[{Shrivastava et~al.(2022)Shrivastava, Larochelle and
  Tarlow}]{shrivastava2022repository}
\bibinfo{author}{Shrivastava, D.}, \bibinfo{author}{Larochelle, H.},
  \bibinfo{author}{Tarlow, D.}, \bibinfo{year}{2022}.
\newblock \bibinfo{title}{Repository-level prompt generation for large language
  models of code}.
\newblock \bibinfo{journal}{arXiv preprint arXiv:2206.12839} .
%Type = Article
\bibitem[{Siddiq et~al.(2023)Siddiq, Santos, Tanvir, Ulfat, Rifat and
  Lopes}]{siddiq2023exploring}
\bibinfo{author}{Siddiq, M.L.}, \bibinfo{author}{Santos, J.},
  \bibinfo{author}{Tanvir, R.H.}, \bibinfo{author}{Ulfat, N.},
  \bibinfo{author}{Rifat, F.A.}, \bibinfo{author}{Lopes, V.C.},
  \bibinfo{year}{2023}.
\newblock \bibinfo{title}{Exploring the effectiveness of large language models
  in generating unit tests}.
\newblock \bibinfo{journal}{arXiv preprint arXiv:2305.00418} .
%Type = Book
\bibitem[{Siddiqui(2021)}]{Siddiqui2021}
\bibinfo{author}{Siddiqui, S.}, \bibinfo{year}{2021}.
\newblock \bibinfo{title}{Learning Test-Driven Development}.
\newblock \bibinfo{publisher}{"O'Reilly"}.
%Type = Article
\bibitem[{Touvron et~al.(2023)Touvron, Martin, Stone, Albert, Almahairi,
  Babaei, Bashlykov, Batra, Bhargava, Bhosale et~al.}]{touvron2023llama}
\bibinfo{author}{Touvron, H.}, \bibinfo{author}{Martin, L.},
  \bibinfo{author}{Stone, K.}, \bibinfo{author}{Albert, P.},
  \bibinfo{author}{Almahairi, A.}, \bibinfo{author}{Babaei, Y.},
  \bibinfo{author}{Bashlykov, N.}, \bibinfo{author}{Batra, S.},
  \bibinfo{author}{Bhargava, P.}, \bibinfo{author}{Bhosale, S.}, et~al.,
  \bibinfo{year}{2023}.
\newblock \bibinfo{title}{Llama 2: Open foundation and fine-tuned chat models}.
\newblock \bibinfo{journal}{arXiv preprint arXiv:2307.09288} .
%Type = Article
\bibitem[{Tufano et~al.(2021)Tufano, Drain, Svyatkovskiy, Deng and
  Sundaresan}]{tufano2021unit}
\bibinfo{author}{Tufano, M.}, \bibinfo{author}{Drain, D.},
  \bibinfo{author}{Svyatkovskiy, A.}, \bibinfo{author}{Deng, S.K.},
  \bibinfo{author}{Sundaresan, N.}, \bibinfo{year}{2021}.
\newblock \bibinfo{title}{Unit test case generation with transformers and focal
  context}.
\newblock \bibinfo{journal}{arXiv preprint arXiv:2009.05617} .
%Type = Inproceedings
\bibitem[{Tufano et~al.(2022)Tufano, Drain, Svyatkovskiy and
  Sundaresan}]{tufano2022generating}
\bibinfo{author}{Tufano, M.}, \bibinfo{author}{Drain, D.},
  \bibinfo{author}{Svyatkovskiy, A.}, \bibinfo{author}{Sundaresan, N.},
  \bibinfo{year}{2022}.
\newblock \bibinfo{title}{Generating accurate assert statements for unit test
  cases using pretrained transformers}, in: \bibinfo{booktitle}{Proceedings of
  the 3rd ACM/IEEE International Conference on Automation of Software Test},
  pp. \bibinfo{pages}{54--64}.
%Type = Inproceedings
\bibitem[{Xie(2006)}]{xie2006augmenting}
\bibinfo{author}{Xie, T.}, \bibinfo{year}{2006}.
\newblock \bibinfo{title}{Augmenting automatically generated unit-test suites
  with regression oracle checking}, in: \bibinfo{booktitle}{ECOOP
  2006--Object-Oriented Programming: 20th European Conference, Nantes, France,
  July 3-7, 2006. Proceedings 20}, \bibinfo{organization}{Springer}. pp.
  \bibinfo{pages}{380--403}.
%Type = Article
\bibitem[{Zhang et~al.(2022)Zhang, Cambronero, Gulwani, Le, Piskac, Soares and
  Verbruggen}]{zhang2022repairing}
\bibinfo{author}{Zhang, J.}, \bibinfo{author}{Cambronero, J.},
  \bibinfo{author}{Gulwani, S.}, \bibinfo{author}{Le, V.},
  \bibinfo{author}{Piskac, R.}, \bibinfo{author}{Soares, G.},
  \bibinfo{author}{Verbruggen, G.}, \bibinfo{year}{2022}.
\newblock \bibinfo{title}{Repairing bugs in python assignments using large
  language models}.
\newblock \bibinfo{journal}{arXiv preprint arXiv:2209.14876} .

\end{thebibliography}


\end{document}